\documentclass[12pt,a4paper]{article}
\pdfoutput=1
\usepackage[utf8]{inputenc}
\usepackage[english]{babel}
\usepackage{extarrows}
\usepackage{amsmath}
\usepackage{amsfonts}
\usepackage{amssymb}
\usepackage{graphicx}
\usepackage{fourier}
\usepackage{physics}

\usepackage{caption}
\usepackage{subcaption}
\usepackage{float}
\usepackage{color}
\usepackage{cancel}
\usepackage{abstract}
\usepackage{appendix}
\usepackage{authblk}
\usepackage{xcolor}
\usepackage{cite}

\numberwithin{equation}{section}

\newcommand\nn{\nonumber}
\newcommand\be{\begin{equation}}
\newcommand\ee{\end{equation}}
\newcommand\ba{\begin{eqnarray}}    
\newcommand\ea{\end{eqnarray}}      
\usepackage[english]{babel}

\title{Real-time methods in JT/SYK holography}

\author[a]{Ra\'ul Arias\footnote{rarias@fisica.unlp.edu.ar}}
\author[a]{Marcelo Botta-Cantcheff\footnote{marcelo.botta.cantcheff@cern.ch}}
\author[b]{Pedro J. Martinez\footnote{pedro.martinez@cab.cnea.gov.ar}}

\affil[a]{Instituto de F\'isica La Plata - CONICET and 

Departamento de F\'isica, Universidad Nacional de La Plata ,

C.C. 67, 1900, La Plata, Argentina}

\affil[b]{Centro At\'omico Bariloche, 

R8402AGP, San Carlos de Bariloche, Argentina}

\date{}
\usepackage[left=2cm,right=2cm,top=2cm,bottom=2cm]{geometry}

\emergencystretch=\maxdimen
\begin{document}

\maketitle
\thispagestyle{empty}
\begin{abstract}
We study the conventional holographic recipes and its real-time extensions in the context of the correspondence between SYK quantum mechanics and JT gravity. We first observe that only closed contours are allowed to have a 2d space-time holographic dual. Thus, in any real-time formulation of the duality, the boundaries of a classical connected geometry are a set of closed curves, parameterized by a complex \emph{closed} time contour as in the Schwinger-Keldysh framework. Thereby, a consistent extension of the standard holographic formulas is proposed, in order to describe the correspondence between gravity and boundary quantum models that include averaging on the coupling constants. We investigate our prescription in different AdS$_{1+1}$ solutions with Schwinger-Keldysh boundary condition, dual to a boundary quantum theory at finite temperature defined on a complex time contour, and consider also classical, asymptotically AdS solutions (wormholes) with two disconnected boundaries.

In doing this, we revisit the so-called factorization problem, and its resolution in conventional holography by virtue of some (non-local) coupling between disconnected boundaries, and we show how in specific contexts, the averaging proposal by-passes the paradox as well, since it induces a similar effective coupling. 
\end{abstract}

\newpage

\pagenumbering{arabic}
\tableofcontents

\section{Introduction}

The AdS/CFT correspondence is a useful tool to study strongly coupled quantum field theories through a gravitational model. In this sense, it allowed to describe properties of many physical situations ranging from hydrodynamics to quark gluon plasma and condensed matter theories \cite{magoo}. In a nutshell, the conjecture states the equivalence between the Hilbert space of the field theory and the Hilbert space of its gravitational dual. Nevertheless a complete map between these Hilbert spaces is not known.
In recent years, a new duality was found between a particular quantum mechanics model (called SYK) and a particular two dimensional gravity theory (JT gravity). It is by exploring this correspondence that an old puzzle of AdS/CFT \cite{WittenFact,MaldaMaoz}, dubbed \textit{the factorization problem}, could be revisited. The puzzle can be presented as follows: the partition function of a set of $n$ non-interacting CFTs should correspond to the product of each theory's partition function, whereas the bulk problem taking the $n$ CFTs as boundary conditions should a priori consider all connected geometries in the expected sum over topologies required for the gravitational path integral. 
A proposal to solve this problem, inspired by this JT/SYK duality, is that the actual holographic dual to the gravitational path integral is not a particular QFT but rather an ensemble of QFTs, which effectively avoids the factorization in the CFT side of the correspondence. 
See for example \cite{Sarosi17,Trunin20} and references within for a review.

More precisely, SYK \cite{SY,K} is a theory (or rather an ensemble of theories) of $N$ Majorana fermions with polynomial all to all interactions of order $q$ via a random coupling constant that has zero mean and Gaussian distribution. This model is invariant under reparametrizations but in the IR this symmetry is spontaneously broken down to $SL(2,\mathbb{R})$. This implies that at the IR fixed point the Goldstone modes of SYK can be described by a one dimensional conformal field theory \cite{Trunin20}. Its holographic dual, called JT gravity \cite{Teitelboim1983,Jackiw1984}, is a two dimensional model that couples a Dilaton scalar field with the geometry. The action can be written in terms of a unique degree of freedom (a reparametrization mode) and is $SL(2,\mathbb{R})$ invariant. Despite being a two dimensional theory it can be obtained from higher dimensional near extremal black hole solutions by dimensional reduction. By analyzing random matrix models it can be seen that the partition function of both theories coincide \cite{Saad2019}.

A particularly puzzling aspect of the JT/SYK duality is related to the averaging process in the QM theory \cite{Sarosi17,Trunin20} 
which is believed to be intimately related to the existence of wormhole geometries.
The standard picture of the holography community is that the correspondence between CFTs and aAdS gravity is a one to one map and many precision holography tests were carried successfully in this front \cite{magoo}. However, for JT computations to match most predictions of SYK, one needed not to consider a single realization of coupled fermions but actually an ensemble of these theories. If the averaging in holography turns out to be mandatory for the higher dimensional scenarios as well, this has some quite strong consequences.
On one hand, this immediately solves the factorization problem raised before, since the averaging effectively make all theories interact with each other \cite{ABS}.
On the other hand, this appears to be a quite unnatural limitation of AdS/CFT (see e.g. \cite{Cotler:2022rud} for discussion) and furthermore it conflicts with numerous non trivial one to one realizations of the correspondence \cite{magoo}.
Whether the averaging is a fundamental piece of AdS/CFT or a peculiarity of the JT/SYK example is still an open debate \cite{AverageDeBoer1,AverageDeBoer2}. 

Even from a purely gravitational perspective, it is by now clear that JT gravity provides a simpler but non trivial model to revisit old and pose new gravitational path integral questions. 
In particular it has been capable of reproducing the conjectured Page curve for the BH entropy \cite{Almheiri19,Prem20}. Many studies and models on entanglement entropy as well as properties of the partitions functions have been exhaustively studied in this context \cite{Saad2019,Stanford19,Witten2020a,Witten2020b}. 

It is important to stress that the study of the map between the QFT and gravity Hilbert spaces requires a clear holographic prescription in real time, such as to understand the possible physical states that can evolve in the system. 
From the foundational works \cite{GKP,W} onward, the Euclidean AdS/CFT prescription was ever-growing, and solid holographic dictionary entries were built \cite{magoo}. For simple enough systems, a Wick rotation of Euclidean results suffices to extrapolate correct real-time predictions. However, the physical interpretation of a Wick rotated Euclidean quantity often requires acute physical intuition on the system's phenomenology and Wick rotated Euclidean correlators are often written as a sum over discrete Matsubara frequencies, whose convergence and/or re-summation is hard to analyze \cite{SvRL}. 

Thus, a purely real-time prescription of holography is still mandatory to fully understand the duality, even if it poses both conceptual and computational challenges.
Skenderis and van Rees (SvR) in \cite{SvRC,SvRL} developed a holographic prescription in real-time holography based in finding gravity duals to generic complex-time Schwinger-Keldysh (SK) contours. The use of complex-time paths to represent real time physical systems is well documented in the QFT literature \cite{Kamenev,Kamenev2,Kamenev3} as well as for AdS/CFT applications, see e.g. \cite{RangamaniBook}.
In short, the SvR prescription sets a gravitational path integral with complex boundary conditions. If the CFT path integral over the SK path is physically well understood and taking holography as a hypothesis, these boundary conditions provide also a well posed and physical variational problem for the gravity path integral. 
Its exact computation is in general out of reach, but often for the cases of interest one can study approximations to this problem. 

The SvR prescription was already proven useful in expanding the map between the dual gravity and QFT Hilbert spaces \cite{us1,CS,Rabideau,Thesis,Belin20,WillyPet}. In particular, inserting boundary sources in the Euclidean segments can be seen to build excited states over the fundamental configuration provided by the saddle point of the sourceless Euclidean manifold. In the large-$N$ limit, these excited states can be seen to become coherent excitations over the vacuum with non-trivial expectation values for the sourced field \cite{us1,CS,Rabideau}, but $1/N$ corrections progressively deform their coherence \cite{us2}. 
Furthermore, the complexified nature of the SvR problem suggest that complex sources for real fields yields relevant physical results. It is the case that coherent states allow for complex parameters corresponding to non trivial expectation values for both the field and its conjugated momentum \cite{Thesis,Belin20,WillyPet}. Thereby, a central piece of this paper is to focus on the study of these prescriptions and methods in the specific arena of two dimensional gravity.

The main point of our work is to provide a systematic presentation of the SvR prescription in the particular scenario of $1+1$ gravity. This well established prescription provides a nice physical interpretation for all observables and quantities computed. In doing so, we find some non-trivial results that we summarize in the end of this section.
We must mention that approaches with similar motivations have risen recently in the literature, see \cite{Garcia22,Loganayagam22,Matsoukas-Roubeas22,Bzowski22}, showing the community interest for this type of analysis and developments. However, the focus of our work is different. Following a well established real-time holographic prescription eliminates ambiguities that rise upon direct analytic extensions of correlators. 

We conclude this introductory section by listing some of the most noticeable results and remarks achieved along the paper:
\begin{itemize}
    \item The holographic GKPW standard dictionary, and its real time extensions (the SvR prescription), is extended to two-dimensional gravity. In the known model of JT/SYK, it involves averaging on the quantum mechanics side. 
    \item Geometric arguments show that this extension can only be done by considering \emph{closed} SK complex paths in the field theory, and the SvR extension of the dictionary to real time is captured by the real time intervals of the path.
    \item The arguments in \cite{ABS} regarding traversable wormholes are revisited and improved in the light of the JT/SYK duality. 
    One can argue that by including a (non local) interaction term between disconnected boundaries in the action could solve the factorization issue. Moreover, it is consistent with the well-established mechanism to construct traversable wormholes \cite{ABS, Gao}. 
    \item A unified equation capturing the holographic real time prescriptions and average ensemble for $b>1$ boundaries is presented in eq. \eqref{SvR-JT-average-manyb}.
    \item The concern of factorization on eq. \eqref{SvR-JT-average-manyb} implies certain restrictions on the distribution of probability of the coupling constants on the boundaries. And in addition, the existence of wormhole as classical gravity solution (dominant saddle point) implies that the duality JT/SYK should be deformed a some non-trivial way.  
    \item Novel real-time extensions of wormholes in JT with sources are studied, where the time-ordered correlation functions are calculated.
\end{itemize}

The paper is organized as follows, in section \ref{Setup} we will review the connection between a SK path and holography using the SvR proposal in dimensions $d+1>2$. In section \ref{Sec:SvR-JT} we will study the specific 2 dimensional case in the JT gravity context dual to a single boundary SK path. We focus on the differences between the $1+1$ and $d+1>2$ scenarios. We present a concrete example of a geometry dual to a single SK path and explore its propagating modes and correlation functions for probe fields.
In section \ref{sec: wh-sols} we focus on the scenario of many SK paths as boundaries and find constraints on the possible form of the ensemble average by demanding consistency with factorization at large $N$. We also present an example of a geometry dual to a pair of boundaries and explore its real time dynamics and correlators.
Finally, we summarize the results obtained and mention some prospects for future work in section \ref{sec:conclusiones}.

\section{SvR in general $d+1>2$ dimensions}
\label{Setup}

In this introductory section we review the Skenderis and van Rees prescription as presented in $d+1>2$ standard AdS/CFT holography. In this context the averaging is ignored. We will improve on this prescription in the upcoming sections by a more suitable approach that captures this ingredient as well as provides a tool to avoid the factorization problem.
First, we review the $d+1>2$ SvR \cite{SvRC,SvRL} construction, providing explicit gravity duals to generic complex-time paths, which we collectively denote Schwinger-Keldysh (SK) paths. These gravity duals can always be split in pure Euclidean or Lorentzian segments. In this set-up we will see that it is natural to interpret asymptotic sources (i.e. external sources insertions in the dual CFT) to the Euclidean segments as excited states of the original theory. In general, these excitations are only tractable as perturbations over a gravitational vacuum and have been covered and studied in the bibliography in recent years \cite{us1,CS,Rabideau,Thesis,Belin20,WillyPet}. Their large $N$ phenomenology look like coherent excitations over the reference vacuum. Real time correlations can be seen to be modified by the insertions of these sources both to leading and subleading orders in $1/N$. Interestingly, even complex sources for real fields have a natural explanation in this set-up.

\subsection{Piece-wise holographic duality}
\label{SvR}

\begin{figure}[t]\centering
\begin{subfigure}{0.49\textwidth}\centering
\includegraphics[width=.9\linewidth] {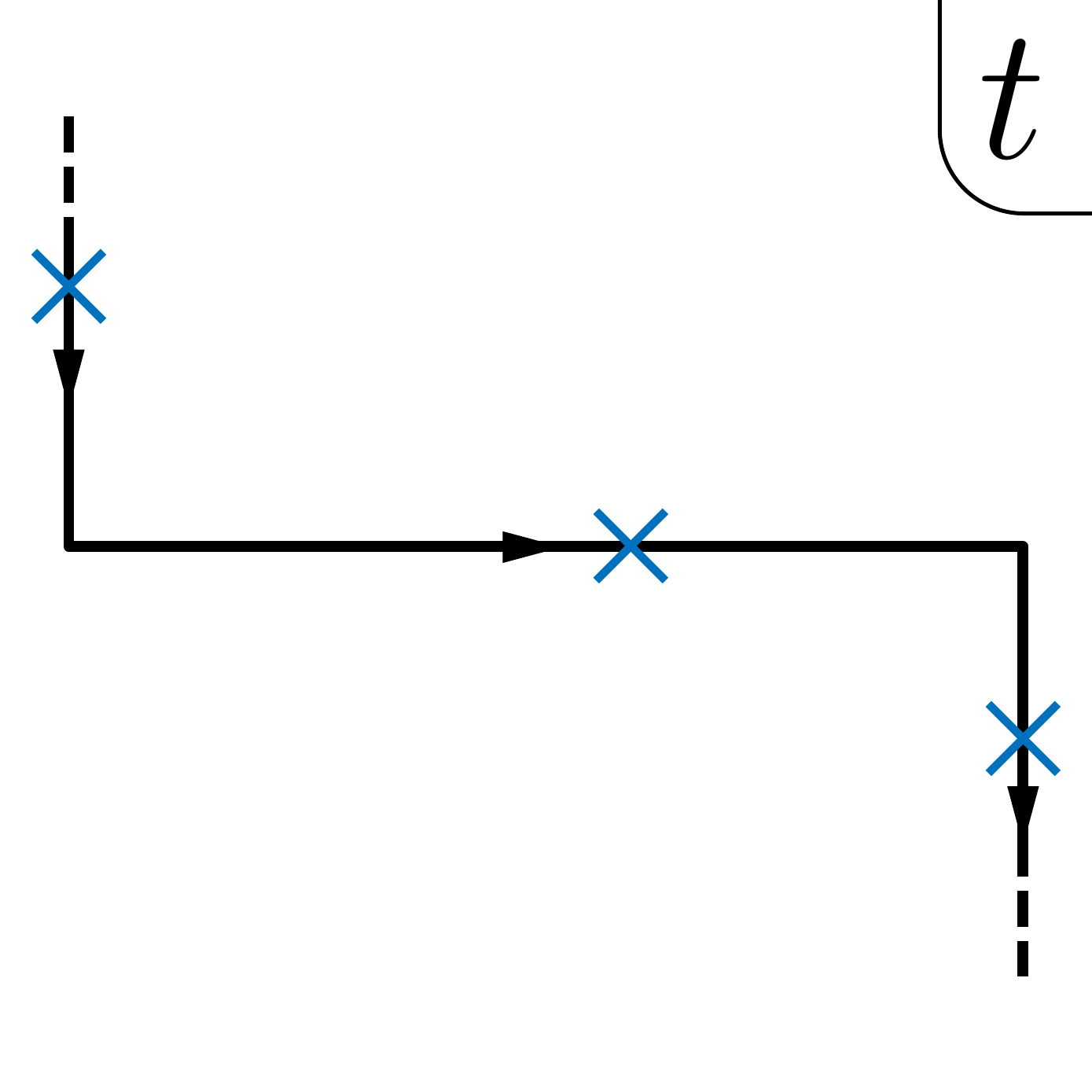}
\caption{}
\end{subfigure}
\begin{subfigure}{0.49\textwidth}\centering
\includegraphics[width=.9\linewidth] {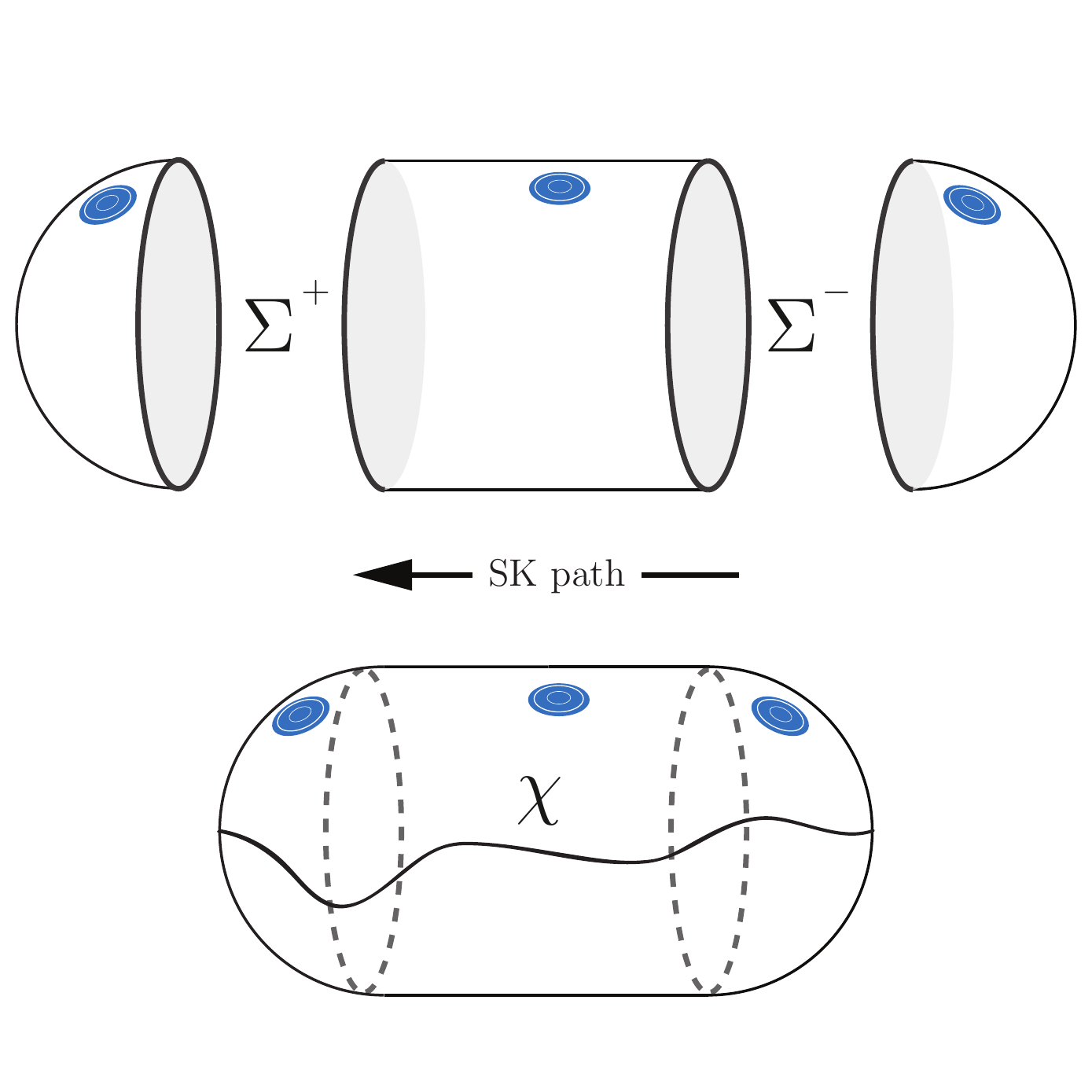}
\caption{}
\end{subfigure}
\caption{(a) In-Out SK path corresponding to a scattering amplitude computation in QFT. Blue crosses represent operator insertions. (b) Bulk dual of the In-Out SK path. The SvR prescription associates half Euclidean AdS spheres and Lorentzian AdS cylinders respectively to the Euclidean and Lorentzian segments of the path on the left. These are $C_1$ glued across $\Sigma^\pm$. The resulting manifold provides a candidate saddle for the gravitational path integral. Asymptotic sources are represented with blue lumps.}
\label{Fig:In-Out}
\end{figure}

To present the SvR formalism, we find convenient to review a very simple example of this prescription, being the case of a QFT scattering process. A more detailed introduction can be found in \cite{SvRL}. The corresponding SK path for our example is presented in Fig. \ref{Fig:In-Out}(a), see \cite{SvRC,us1}. 
The dynamics take place in the Lorentzian segment and (time-ordered) $n$-point functions are computed via external sources inserted there. The initial and final states are defined by the Euclidean regions and their asymptotic sources, the vacuum state being defined as the Euclidean path integral with all sources in the Euclidean segments turned off. The bulk dual is shown in Fig. \ref{Fig:In-Out}(b): Euclidean half-sphere sections and Lorentzian AdS cylinders are assigned to each Euclidean and Lorentzian SK segment respectively and ${C}^1$-glued across $\Sigma^{\pm}$, providing a candidate saddle for the full path integral. Classical bulk fields configuration $\chi$ are fully determined in terms of its asymptotic Lorentzian $\chi_L$ and Euclidean $\chi_{\pm}$ boundary conditions. We write the holographic relation in this set-up as,
\begin{align}\label{eqSvR0}
Z^{CFT_d}_{\chi_-\to \chi_+}[\chi_L]=\langle \chi_+|e^{-i\int {\cal O}\chi_L}|\chi_-\rangle\equiv Z^{AdS_{d+1}}_{\chi_-\to \chi_+}[\chi|_{\partial}=\chi_L],
\end{align}
where $Z^{CFT_d}_{\chi_-\to \chi_+}[\chi_L]$ is the CFT$_d$ generating function of correlators between the states defined by $|\chi_-\rangle \equiv e^{-\int {\cal O}\chi_-}|0\rangle$ and the final state $\langle\chi_+|\equiv(|\chi_+\rangle)^\dagger$, see \cite{Jackiw,us1}. Real time correlators are obtained by differentiation wrt $\chi_L$. On the gravity side, we have 
\begin{align}\label{eqSvR}
Z^{AdS_{d+1}}_{\chi_-\to \chi_+}[\chi|_{\partial}=\chi_L]=  \left(\int_{\chi_-}{\cal D}\chi \; e^{-I_E}\right) \left(\int_{\chi_L}{\cal D}\chi \;e^{- i I_L}\right) \left(\int_{\chi_+}{\cal D}\chi\;e^{-I_E}\right)
\end{align}
with $I_{L/E}$ are the Lorentzian/Euclidean corresponding local gravity actions on each segment. The smooth Israel gluing at $\Sigma^\pm$ that links the factors in the rhs is left implicit.
This expression suggests a \emph{piece-wise holographic recipe} with intervals in which the dynamics take place (Lorentzian) and intervals in which information on the system's state is given (Euclidean).

\subsection{Holographic excited states}
\label{Sec:HES}

After introducing the SvR formalism and its piece-wise holographic prescription, we pay special attention to the Euclidean segment defining the initial state, first factor on the rhs of eq. \eqref{eqSvR}.
Deformations on this region (keeping the rest of the manifold including the $\Sigma^-$ surface fixed) can be thought as providing a different initial state to the real time evolution problem. When the deformation is given by turning on CFT external sources, we collectively call these excitations holographic excited states \cite{us1,CS,us2,us3,us4,us5,Rabideau}. 
Its bulk wavefunction is obtained via a Hartle-Hawking Euclidean path integral with non-trivial asymptotic boundary conditions. On the CFT side, their interpretation is also natural as excited states due to operator insertions in the Euclidean past. The sources turning off softly near the $\Sigma$ gluing regions guarantees that the Hilbert space at the meeting point is still given by the vacuum (reference) Hilbert space and allows for the ``excited state'' nomenclature to make sense.
To be explicit, in \eqref{eqSvR} we have implicitly defined a state $|\chi_-\rangle$ on each side of the dual pair as
\begin{equation}
\label{exc-state}
\text{\sf CFT:}~~~\langle A_{\partial\Sigma}|\chi_-\rangle\equiv  \int_{A_{\partial\Sigma}}{\cal D} A \; e^{-I_{CFT}-\int \chi_-\; {\cal O}} \qquad\Longleftrightarrow\qquad \langle \chi_\Sigma|\chi_-\rangle=\int_{\chi_\Sigma;\chi_-} {\cal D}\chi \; e^{-I_E}
~~~:\,\text{\sf AdS}\end{equation}
Here $A$ denotes collectively the fundamental CFT fields and $\chi$ bulk fields dual to CFT primaries ${\cal O}$. The integration on the Euclidean gravitational path integral over the metric is left implicit. We have denoted $A_{\partial\Sigma}$ and $\chi_\Sigma$ to the field-configuration basis at $t=\tau=0$. Notice that this does not conflict with the standard Euclidean intuition that external CFT sources $\chi$ translate into boundary conditions for bulk fields $\chi$ under the holographic map.

As we mentioned above, most of the studies on these states were done in a perturbative way, i.e.
\begin{equation}
    |\chi_-\rangle={\cal P}\{ e^{-\int {\cal O}\chi_-} \}|0\rangle \sim \left(1-\int {\cal O}\chi_-+\frac 12 \int {\cal P} \{{\cal O} {\cal O}\}\;\chi_- \;\chi_-+\dots\right)|0\rangle\;.
\label{CFTex}    
\end{equation}
Excitations of this sort can be built over any state of the theory but in general one is often interested in reference states that have a known bulk dual in the semi-classical limit. Over these, states \eqref{CFTex} will also have a semi-classical bulk dual as long as backreaction is under control. To clarify, the reference state can be the vacuum as in the scattering example above, a finite temperature state as it will be the case for most of this work, or any other state accessible through an Euclidean path integral. 

In the large $N$ limit, single trace operators $\cal O$ become generalized free fields. This is explicitly realized by the BDHM  \cite{BDHM} prescription, which for a scalar operator ${\cal O}$ of conformal dimension $\Delta$ can be written as
\begin{equation}\label{BDHM}
    {\cal O}\equiv (2\Delta-d ) \lim_{r\to\infty} r^{\Delta} \chi
\end{equation}
where $r\to\infty$ is taken to be the limit to the asymptotic boundary and $r^{\Delta}$ allows to retain the leading $r^{-\Delta}$ contribution from the canonically quantized field $\chi$ of mass $m^2=\Delta(\Delta-d)$. We should stress the operator character of this equation: for weakly interacting bulk matter fields it implies that ${\cal O}$ has a nice representation in terms of the bulk ladder operators.
Then, each term in the series in eq. \eqref{CFTex} has a natural $n$-particle state interpretation. In the strict $N\to\infty$ limit the bulk matter fields become free and ${\cal O}$ becomes linear in the bulk ladder operators. The state then becomes coherent \cite{us1} with $1/N$ corrections gradually deforming their coherence property \cite{us2}.  
Once the connection between \eqref{exc-state} and coherent states is made one can import some intuition built from the latter to the former. For our purposes, a particularly useful intuition is to allow the sources $\chi_-$ to become complex valued and assign the real and imaginary parts of $\chi_-$ its standard physical interpretation that relates it to the $\chi$ and its conjugated momentum $\Pi_\chi$ expectation values \cite{Thesis,Belin20} 
\begin{equation}
\langle\chi_-|\;\chi\;|\chi_-\rangle=\int_{\chi_-}{\cal D}\chi \;\chi\; e^{-I_E}\propto \text{Re}[\chi_-], 
\qquad\qquad
\langle\chi_-|\;\Pi_\chi\;|\chi_-\rangle=\int_{\chi_-}{\cal D}\chi \;\Pi_\chi\; e^{-I_E}\propto \text{Im}[\chi_-]\;.
\label{bcEOM}
\end{equation}
where $\langle\chi_-|$ is built using Euclidean conjugation \cite{Jackiw}, i.e. conjugation and time reflection on $\chi_-$. This guarantees that both results are real even for $\chi_-\in\mathbb{C}$.

Note that we are taking a complex source $\chi_-\in\mathbb{C}$ for a real bulk scalar field $\chi$ which seems to conflict with the counting of degrees of freedom.
What happens for a complex source is that this curve no longer lies on the real axis, as did in the purely Lorentzian set-up.

The main take away from this section is that SvR provides a real time holographic prescription in $d+1$ holography in which, for example, to study real time correlations and build excited states over a reference state and allows for both real and imaginary sources for real fields. Specifically, imaginary sources for an operator $\cal O$ in the CFT translate in the bulk to a non-trivial expectation value for the conjugate momentum $\Pi_\chi$. 
However, notice that this prescription is unable clarify the factorization problem as presented in the introduction. In the upcoming sections, we aim at improving this real time holographic prescription in the 1+1 dimensional set-up to account for this.

\section{Holographic recipes in 1+1 spacetime dimensions}
\label{Sec:SvR-JT}

The main question addressed in this paper is if conventional holography holds in the context of Jackiw-Teitelboim gravity in the same way as the standard holographic prescriptions as GKPW, or its generalization to real-time: the SvR formula eq. \eqref{eqSvR0}. The validity of the holographic map between operators eq. \eqref{BDHM} shall also be studied in this context \cite{BDHM}.
In this section we will study how to understand the SvR construction in two dimensions, showing that open SK paths does not have a corresponding gravity dual state. Then we will add a probe scalar field $\chi$ to the JT action to compute correlators of quantum fields living on the boundary theory and will mention some features of holographic excited states in this framework. 

\subsection{The SvR formalism: No open paths in 2d-holography }
\label{Sec:<SvR>}

First, we are going to see that the SvR formula for one complex closed boundary can be formulated in a conventional way simply by taking an  evolution operator $U :{\cal H} \to  {\cal H}$, which is a path ordered (Hermitean) operator, valued on a \emph{closed} oriented curve ${\cal C}$ embedded in the complex plane $\mathbb{C}$ as 
\begin{equation}\label{ZCFT}
Z_{QFT} = \text{Tr} \, \;U[\kappa, {\cal C}] \qquad\qquad U[\kappa, {\cal C}] \equiv {\cal P} \,e^{-i\oint_{{\cal C}_{}} d\theta\; (H + {\cal O} \,\kappa(\theta))},\qquad\theta \in \mathbb{C},
\end{equation}
where $U[\kappa,{\cal C}]$ is the evolution operator for a boundary Hamiltonian deformed by a source $\kappa(\theta)$. This operator acts on the Hilbert space ${\cal H}$ of the suitable $0+1$-dimensional quantum field theory. 

Thereby, the conventional SvR formula takes the simple form
\begin{equation}\label{SvR-JT}
\text{Tr} \, \;U [\kappa, {\cal C}] = Z_{grav} (\kappa) \qquad\qquad \partial M \equiv {\cal C}\qquad \kappa = \chi|_{\cal C}
\end{equation}
where $\kappa$ are Dirichlet boundary conditions to probe matter fields $\chi$
defined on the (locally AdS) spacetime $M$, whose boundary is ${\cal C}$. This is the formulation of \cite{us3,us4} in higher dimensions. 
In particular, this formula also applies to the purely Euclidean set up \cite{GKP,W}, where the SK closed contour ${\cal C}$ corresponds to the Euclidean circle of length $\beta$, and the dual geometry is the Euclidean AdS disc.

Notice that, eq. \eqref{SvR-JT} can only be formulated for \emph{closed} complex boundaries. In other words: \emph{there is no states/evolution} on open complex boundaries, which are dual to some classical geometry.
By strictly geometric arguments one can show that the SvR(GKPW) recipe, in the context of two dimensional gravity, can only be formulated  for closed boundary curves; in fact, in order to have a well defined rhs of \eqref{SvR-JT}, in the semiclassical limit, it has two dimensional classical geometries $M$ as saddles of a well posed problem and it requires that the boundary is given by closed curves\footnote{Actually, the argument does not requires the semiclassical limit or saddle geometries; in fact, already the JT partition function is well defined as a sum over Riemannian two dimensional manifolds whose boundary is restricted to be $\sqcup^b {\cal C}$ }. In other words, in the simpler purely Euclidean version of eq. \eqref{SvR-JT}, $M$ is a general Riemann surface with $b$ boundaries and $g$ handles where each boundary is equivalent to a circle, i.e: $\partial M = S^1 \sqcup \dots \sqcup S^1$, with circumference lengths $\beta_1 \dots \beta_b$ respectively.

\subsubsection{A real-time prescription for JT/SYK: SvR with averaging}
\label{Sec:SYK-SvR}

The Sachdev-Ye-Kitaev (SYK) model \cite{SY,K} is a quantum mechanical theory for $N\gg1$ Majorana fermions with an all-to-all interaction. The action is written in Euclidean time $\tau$ as
\be
I_{SYK}=\,\,\int d\tau\left(\frac12\sum_{i=1}^N\psi_i(\tau)\dot\psi_i(\tau)+\frac{1}{4!}\sum_{i,j,k,l=1}^N J_{ijkl}\psi_i(\tau)\psi_j(\tau)\psi_k(\tau)\psi_l(\tau)\right).\label{SYKaction}
\ee
The Majorana fermions are Hermitian and satisfy the canonical anticommutation relation $\{\psi_i,\psi_j\}=\delta_{ij},\,\,\,\, i,j=1,\ldots,N$. The coupling $J_{ijkl}$ are taken randomly and in an independent way from a Gaussian distribution function with the probability density
\be
G_\sigma(J_{ijkl})\equiv e^{-\frac{J_{ijkl}^2}{2\sigma^2}} \qquad \text{where} \qquad\sigma^2 \equiv \frac{3!\sigma^2_0}{N^3}\label{gaussian}
\ee
In this distribution the mean coupling and the squared variance are defined
$$
\mu\equiv \langle J_{ijkl} \rangle\equiv  \int dJ\, G_\sigma(J) \,  J_{ijkl}=0, \qquad \;\qquad   \langle J_{ijkl}^2 \rangle \equiv\int dJ\, G_\sigma (J) \, J_{ijkl}^2 = \frac{3!\sigma^2_0}{N^3} \, .
$$ 
Note that $\sigma$ is a constant with mass dimensions.

Lastly, note the action \eqref{SYKaction} can be written with a more general $q$-interaction term with coupling constant $J_{i_1 \dots i_q}$ with $q$ indices, although for simplicity, we focus on the model $q=4$ in \eqref{SYKaction} and in the rest of the paper.

Since SYK with random interactions is believed to be the dual of JT gravity \cite{Trunin20,Saad2019}, one should generalize the prescription of the previous Section to describe averaged holographic theories. This might be done by correcting the formula \eqref{SvR-JT} on the left hand side as
\begin{equation}\label{SvR-averaged}
\text{Tr} \, \;U [J, \kappa, {\cal C}]\qquad\to    \qquad \langle\text{Tr} \, \;U [\kappa, {\cal C}]\rangle \equiv \int dJ\, G(J) \, \text{Tr} \, \;U [J, \kappa, {\cal C}]
\end{equation}
where $U[J]$ denotes the evolution operator of the boundary field theory, generated by the Hamiltonian $H[J; \psi, \dot \psi]$.
In general, $J$ denotes the set of parameters (coupling constants) labeling a family of boundary quantum theories. The averaging is weighted by the function $G(J)$ which, in particular 
in the SYK model, is taken to be the Gaussian distribution \eqref{gaussian} and $J\equiv J_{ijkl}$, with $i,j,k,l=1,\dots,N$

Thus, our proposal for the holographic map is similar to the SvR formula, with the averaging on the coupling constants of the boundary quantum theory
\begin{equation}\label{SvR-JT-average}
 \langle\,\text{Tr}\, U\left[\kappa ; \,{\cal C}\right] \rangle = Z_{JT} (\kappa), \;
\end{equation}
where ${\cal C}$ is closed, by virtue of the arguments of the previous subsection, and 
the left hand side stands for an average on $\text{Tr}\, U$, which consists of a path integral \eqref{ZCFT} on that curve. 
The  generalization of this formula  to $b$ closed boundaries ${\cal C}_1, \dots, {\cal C}_b$ is straightforward, and shall be presented in Sec 4 for clearness. We shall see below that this prescription has immediate consequences in the structure, and interpretation of the states, as well as their dynamics.

The present prescription, and its generalization to many boundaries, are  consistent with the \emph{annealed disorder} framework\footnote{This approach is called ``annealed'', in contrast with the so called ``quenched'' disorder where the averaging is taken on the free energy. In the annealed disorder the averaging is taken directly on the partition function, see e.g. \cite{Sarosi17}.} \cite{Sarosi17,Sachdev:2015efa}.
They will be tested in the forthcoming sections, in the specific arena of known JT (and JT$+$scalar field) solutions.

\subsection{Thermal (random) states as holographic dual of 2d spacetimes}
\label{states-1bdy}

The aim of this part is to describe the holographic states, namely, states of the boundary quantum theory that are dual to some classical solution of the JT model, with or without matter fields. 

As shown above, the corresponding field theory path integrals, lhs of \eqref{SvR-JT}, are defined on closed paths ${\cal C}$ which cannot be parameterized by only one real-time interval $t \in [0, T] \subset \mathbb{R} \mapsto {\cal C}$, because it would be a closed time-like curve where the evolution operator is defined on. Instead, all SK path  ${\cal C}$ necessarily has Euclidean intervals $E_\alpha$ of length  $l_\alpha$ respectively, and the total length is  $\beta \equiv\sum_\alpha l_\alpha$.

As explained before, each interval is endowed with an Euclidean evolution operator 
$$U(J, E_\alpha) : {\cal H}_N \to {\cal H}_N\;.$$ 
The microscopic theory is SYK model, whose dynamics characterized by mean $\mu =0$, and arbitrary parameters $q, \sigma_0, N$. The random variables are $2^{N/2} \times 2^{N/2}$ real coupling constants denoted by $J$. This operator unambiguously characterizes a  \emph{microscopic} pure state, that can be represented as a ket $\Psi(J, E_\alpha)$ in the tensor product of two-copies of ${\cal H}_N$ \footnote{ This can be easily seen by representing the matrix elements of $U(J, l_\alpha)$ as a path integral with fixed fields ($\psi_i$) at $\tau$ and $\tau + l_\alpha$. Then it is nothing but the wave functional $\Psi[\psi_i(\tau), \psi_i(\tau + l_\alpha)]$ .}.
This is similar to the standard purification technique in the TFD formalism, see e.g. \cite{us3, us4} for more details.

Therefore, we can naturally define the density operator by considering the (path ordered) union of the Euclidean intervals
\be\label{rhogen}
\rho_\beta \left(J \,,\, \cup_{\alpha} E_\alpha \right) \equiv\,  \;{\cal T} \,\prod_\alpha \; U(J, E_\alpha) = 
U(J, E_1)\, U(J, E_2)\, U(J, E_3)\, \dots =U_\beta\left(J \,, \; \cup_{\alpha} E_\alpha \right) \;, \ee
where taking the trace $\rho_\beta\, \mapsto \,\text{Tr}\, \rho_\beta \,\equiv Z(J, \beta)$ to compute the probability of the $J$-model \eqref{SYKaction}, geometrically represents to close the curve 
$$ \bigcup_{\alpha}\, E_\alpha\, \mapsto\, S^1 \;.$$
In order to keep the probabilistic interpretation of this operator, we must demand hermiticity
\be
\label{rho-hermitic}\rho^{}_\beta = \rho_\beta^\dagger \;.
\ee
In addition, one can define the \emph{effective}, or macroscopic, density operator by integrating out the random variable $J$ in this expression, namely
\be\label{rhogen-macro}
\langle \rho_\beta \rangle \left( \cup_{\alpha} E_\alpha \right) \equiv\,  \int\, dJ \; G_\sigma(J) \;\;
\rho_\beta \left(J\, \,,\, \cup_{\alpha} E_\alpha \right)  \; .\ee
This object is what, in averaged holography, should more naturally be considered the corresponding dual to a classical Euclidean space-time.
Analogously to the standard SK closed-path contour in the complex plane it is \emph{periodic} in the imaginary total time, by virtue of the identification of the endpoints imposed by the trace.
By composing this fact with the statement that all two-dimensional geometry have closed boundaries ${\cal C}$ (which are circles $S^1$ with possible insertions of segments parameterized by real time intervals), we can remarkably conclude that
\emph{the states that are dual to some (JT) geometry, are thermal}, i.e: they can be described by a density operator \eqref{rhogen} with an associated inverse temperature $\beta$, in the same sense that the conventional closed SK path.

\subsubsection{TFD purification}\label{TFDpurification}

Consider now just two equal Euclidean intervals $l_\alpha \equiv \beta/2, \; \alpha =1,2$ on only one disconnected boundary ${\cal C}$, such that
$$
U_2(J\,; \,(0, \beta/2)) = \left( U_1(J\,;\, ( -\beta/2, 0))\right)^\dagger\;,
$$ 
which guarantees \eqref{rho-hermitic}.
In other words, this means that the corresponding path integral on the intervals $E_1 \equiv (0, \beta/2) \;,\; E_2 \equiv E_1^\dagger = ( -\beta/2, 0)\;$, are reflected into each other \cite{us3, us4}. 
The (microscopic) thermal density operator
\eqref{rhogen} is $$ \rho_\beta(J)= U(J\,; \,(-\beta/2, \beta/2)) = \, U_1 (J\,;\, ( -\beta/2, 0)) \, U_2 (J\,; \,(0, \beta/2))\;
= \, U_1 (J\,;\, ( -\beta/2, 0)) \,  \left( U_1(J\,;\, ( -\beta/2, 0))\right)^\dagger.$$
In this way, the operator $U_1(J\,; \,(0, \beta/2))$ itself can be identified with the TFD state, and its thermal excitations can be systematically obtained  by taking a non trivial function $\kappa_1(\tau) \neq 0$ on $E_1$ \cite{us4}.

For instance in the SYK theory for the Grassmannian field $\psi(t)$, the fundamental (TFD) state can be described by the wave functional 
\be
  \Psi(J, \psi(0), \psi(-\beta/2)) \equiv \langle \psi(-\beta/2) | U_1(J) | \psi(0) \rangle \\
    = \int {\cal D}\psi \;  e^{-\int_{\beta/2}^0 d\tau \,\,{\cal L}_{SYK}[\psi, J]} 
\ee
which for an arbitrary constant $J$, is a well defined path integral with fixed arbitrary data $\psi(0), \psi(-\beta/2)$ on the endpoints of $E_1$. Therefore, by following the methods of \cite{Sarosi17} to integrate out $J$, one can compute the components/matrix elements of the (macroscopic) fundamental state \eqref{rhogen-macro}
\be
 \langle \rho_\beta \rangle \equiv \int\, dJ \; G_\sigma(J) \;\; \langle \psi(-\beta/2) | U(J) | \psi(\beta/2) \rangle \\
    = \int {\cal D}\psi \;  e^{-\int_{-\beta/2}^{\beta/2} d\tau \,\frac{1}{2}\, \psi \;\dot \psi  \; +\;\frac{3\sigma_0^2}{N^3}\int^{\beta/2}_{-\beta/2} d\tau\int^{\beta/2}_{-\beta/2} d\tau' (\psi(\tau)\cdot\psi(\tau'))^4} \;.
\ee
This is a path integral that only depends on the initial/final $\psi(\beta/2), \psi(-\beta/2)$ arbitrarily chosen on the endpoints of $E_1 \cup E_2$. It expresses what we call effective, or macroscopic thermal state in SYK. This state should encode most of the features of the (Euclidean) black hole solution of JT gravity. 

\subsection{Canonical quantization and BDHM correspondence in JT/SYK: excited states}

Let us consider $M$ with just one connected boundary curve $ \partial M \equiv {\cal C} $ with
a set of real-time intervals in the same curve, parameterized by $t_\alpha$
where $\alpha$ labels the real-time segments.
 
The simplest scenario to study probe excitations over a geometry is to solve a Klein-Gordon equation 
for a (non back-reacting) scalar field $\chi$ of mass $m^2=\Delta(\Delta-1)$, which is dual to a single trace (scalar) local operator ${\cal O}(t)$ of conformal weight $\Delta$, which typically is only a functional of the fundamental fields of the boundary theory, e.g. $\psi_i$ in SYK. Nevertheless, further generalizations are also possible where it also depends on a random coupling constants (e.g \cite{Garcia20}), that will be considered below. 

As indicated in the recipe \eqref{SvR-JT-average}, the boundary value of $\chi$ is what sources the ${\cal{O}}$ insertions on ${\cal C}$, and allows to compute the generating function of the boundary quantum theory (lhs of \eqref{SvR-JT-average}). Furthermore, in\cite{us1,us4} it was shown that, by quantizing $\chi$ canonically, and applying the BDHM rules \cite{BDHM} one can systematically construct the excited states, which generally correspond to coherent states in the large-$N$ bulk Hilbert space. The aim of this short section is precisely to reproduce this in the JT/SYK context.

The vacuum solution of JT gravity with only one connected boundary ${\cal C}$ consists of exact AdS spacetime, with piece-wise signature as corresponds to the different imaginary/real-time segments of ${\cal C}$, as discussed in Sec. \ref{SvR}.
Consider an appropriate number of locally real time AdS pieces (charts) $ M_\alpha = \{ (t_\alpha, r_\alpha)\}$ covering a time-independent spacetime $M \equiv \{(g_{\mu\nu}(r_\alpha), \Phi(r_\alpha))$. 
For a system of coordinates covering  AdS$_{1+1}$ there are a complete set of normalizable solutions of the KG equation on such geometry of frequencies $\omega_k$ labeled by $k$. 
Thereby, the general solution for $\chi$ can be expressed as
 \be \label{Qcanonical}
\chi \;=\;  \chi_0\; + \; \sum_\alpha \, \, \sum_{k}^{} \, \Theta _\alpha \, \left(a^{}_{\alpha, k} e^{i\omega_k t_\alpha} \, f_k (r_\alpha )\, \pm \, a^\dagger_{\alpha, k}  e^{-i \omega_k t_\alpha }\, f^*(r_\alpha)\right) , \qquad [a^{}_{\alpha, k} , a^\dagger_{\alpha, l}] = \delta_{kl} .
\ee
 where $\Theta_\alpha$ is the Heaviside distribution with support on $M_\alpha$ and 
$ \chi_0 \equiv \sum_\alpha \, \, \Theta _\alpha \, \chi_0(r_\alpha, t_\alpha)$ stands for the non-backreacting classical part of the solution that 
fulfills the non vanishing (Dirichlet) boundary conditions $$\partial M \equiv {\cal C}\qquad \kappa = \chi_0|_{{\cal C}} \; .$$

On the other hand, the (linear) fluctuations are canonically quantized by promoting the coefficients of the general solution to operators as in the rhs of \eqref{Qcanonical}.
In order to have just one probe degree of freedom, the field $\chi$ can be taken taken hermitian (+) or anti-hermitian  (-).
The second choice describes certain modified JT gravity \cite{Garcia20}, which admits interesting wormholes solutions ,e.g. see Sec \ref{Sec:Orandom1b}.

Summarizing, the BDHM recipe allows to connect this field to the operators inserted on the boundary theory
\be\label{BDHM2}
{\cal O}(t_\alpha ) \equiv (2\Delta-1) \lim_{r_\alpha \to \infty} \;  r_\alpha^{\Delta}\,\left( \chi-\chi_0 \right)  \sim \sum_{k}^{} \, \, \left( N_{\alpha, k}\, a^{}_{\alpha, k} e^{-i\omega_k t_\alpha} \, \, \pm \, N_{\alpha, k}^* \,a^\dagger_{\alpha, k}  e^{i\omega_k t_\alpha }\,\right)
\ee
where the radial coordinate $r_\alpha \to \infty$ locally describes each asymptotic region of $\partial M_\alpha$, so 
the relative coefficients in the combination above are given by $N_{\alpha, k} \equiv  \lim_{r_\alpha \to \infty} r_\alpha^{\Delta} \; f_{\alpha, k} (r)$ and $N_{\alpha, k}^*  \equiv  \lim_{r_\alpha \to \infty} \; f_{\alpha, k}^*(r)$.
We show an explicit realization of this construction in Sec \ref{CorrSec3}.

This relation is crucial to describe the excited states around the (thermal) vacuum described in the previous section. Following the same procedure of ref \cite{us4}, plugging this into 
 \eqref{ZCFT}, and using the definition \eqref{rhogen}, one find that these excitations consists of \emph{thermal coherent states} in the bulk Hilbert space. Schematically
$$  \rho_\beta (\kappa) \sim \rho^{(JT)}_\beta \;\otimes \;\left( :\,  e^{-\frac{\beta}{2} \,\hat a_k^\dagger \hat a_k + \kappa_k \, \hat a_k^\dagger + \kappa_k^* \, \hat a_k } \, : \right)\;$$
for $\kappa$ small enough such that the back reaction is negligible and so, the JT gravity sector factorizes from the $\chi$-excitations. 
The $\alpha$-indices were omitted to simplify the notation, $\hat a_k, \hat a_k^\dagger$ denote the appropriate Bogoliubov's transformation (a linear combination of $a_k, \hat a_k^\dagger$) that diagonalizes the Hamiltonian, and $\kappa^{}_k , \kappa_k^* $ are related to the (Wick-rotated) Fourier transform of the sources $\kappa(\tau)$ on the imaginary-time segment $(0, \beta)$, see \cite{us4}.

It is worth emphasizing that the standard holographic formula \eqref{BDHM} \cite{BDHM}  implicitly assumes that the local operators in the lhs of \eqref{BDHM2} belong to the SYK theory, i.e. they only depends on the fields of the theory and its time derivatives, i.e. ${\cal{O}} = {\cal{O}}(\psi_i , \dot \psi_i , \ddot \psi_i, \dots )$. 
 We will show in a subsequent section that the inclusion of bulk fields with non-trivial back reaction (sJT) might require a significant deformation of the randomly coupled SYK model, and the standard BDHM formula should be properly modified involving averaging. For instance, in presence of operators depending on the (random) coupling constants (e. g. see the explicit example of Sec \ref{Sec:Rigidity-Wh-G2}) , the simplest modification of \eqref{BDHM2}, which is consistent with \eqref{SvR-JT-average}, can be expressed as 
\be\label{BDHM2-ave}
{\cal O}(t) \equiv  \int d J\, d M\,  G(J) \, G(M) \, \;   {\cal O}(J, M, t) \, \,\, e^{-I_{}[\psi, J, M, {\cal C}]}    = (2\Delta-1)\lim_{r\to \infty} \;  r^{\Delta}\,(\chi(r, t) - \chi_0 )\,\,,
\ee
The notation stands for an averaged \emph{operator},
where $M$ denotes all the extra randomly distributed constants, introduced in the model through the operator ${\cal O}$, see example below. 
This formula is also going to be tested among the forthcoming examples. 

\subsubsection{SYK with random deformations}
\label{Sec:Orandom1b}

Let us illustrate a situation described above with the simplest toy example. An interesting model inspired in  \cite{Garcia20} is described by the SYK Hamiltonian
\be
H[J] \equiv  \frac{1}{2}(\psi_i\,  \cdot \, \dot \psi^i) + J_{ijkl} \psi^i \psi^j\psi^k\psi^l  ,
\ee
is the unperturbed microscopic Hamiltonian on a single $0+1$d boundary. 
The type of perturbation proposed in ref \cite{Garcia20} is
 $$ V[M, t] \equiv \, i  \; \kappa(t) \,{\cal O}[M] \, ,
 $$ 
 where the operator generating the deformation is defined as
 \be  \label{OpeM}
 {\cal O}[M] \equiv M_{ijkl} \psi^i \psi^j\psi^k\psi^l \,,
 \ee 
and the perturbation is sourced by an arbitrary function $\kappa$ along ${\cal C}$. This scenario is slightly different from the general framework where the formula \eqref{BDHM} applies, since the operator that generates the perturbation is, itself, associated to an (independent) random coupling $M$, independent from $J$'s (see e. g. \cite{Garcia20} and references therein). Essentially, this is nothing but a standard SYK theory with $q=4$, by perturbing it with a purely imaginary term
 $$J\; \mapsto\,\,j\equiv J + i \kappa M\; ,$$
 where both $J, M$ denote real $2^{N/2} \times 2^{N/2}$ matrices.
Thereby, the model is SYK with \emph{complex} random coupling
  \be\label{U-averaged}  \langle U(j  , {\cal C}) \rangle \equiv  \; \frac{\pi \sigma^2}{2} \int dJ_{ijkl}\, dM_{ijkl} \, \, G_\sigma(J_{ijkl})\,\, G_\sigma(M_{ijkl}) 
\;  {\cal T} \;e^{-\int_{{\cal C}} d\theta \; \left( \frac{1}{2}(\psi\, \cdot \,\dot \psi) + \frac{1}{4!}\sum_{i,j,k,l=1}^N \, j_{ijkl} \psi^i \psi^j\psi^k\psi^l \right)}, \ee
where the double bracket represents the averaging on the two independent random parameters $J, M$. By integrating out $M$, we recover the averaged evolution operator
\be \label{estadoG2} \langle U(J ) \rangle \equiv \int dJ_{ijkl}\,  \, \, G_\sigma (J_{ijkl})\,\,
\;  {\cal T} \,e^{-\int_{{\cal C}} d\theta \; \left( \frac{1}{2}(\psi\, \cdot \, \dot \psi) + \frac{1}{4!}\sum_{i,j,k,l=1}^N \, J_{ijkl} \psi^i \psi^j\psi^k\psi^l  + \tilde V(\theta)\right)}
\ee
which is the standard SYK model, but corrected with an \emph{effective} non-local potential defined as 
\be\label{non-localO}
\tilde V (\theta) \, \equiv \; - 
\; \frac{\sigma^2}{2}\;\;\int_{{\cal C}} d\theta'\, \kappa(\theta) \kappa (\theta') (\psi(\theta)\cdot\psi(\theta' ))^4 \;.
\ee

Notice that, according to the considerations above, the corresponding bulk dual could be modelled by a purely imaginary extra free scalar field $\chi$ on a fixed JT background geometry \cite{Garcia20}. The solutions are similar to the studied in section \ref{CorrSec3} but with purely imaginary boundary values
$$ \lim_{r\to\infty}r^{-(\Delta-1)} \,\chi (r,t) = i\kappa(t) \, ,$$ 
and the canonical quantization of its fluctuations is to be implemented according to \eqref{Qcanonical}, with the \emph{minus} choice. Thereby, the suitable holographic extrapolation recipe may be realized by the formula \eqref{BDHM2-ave}.
In fact notice that by taking a derivative of the operator \eqref{U-averaged} with respect to $\kappa(t)$, we obtain the lhs of that formula \eqref{BDHM2-ave}, while the counterpart in the bulk, is given by the quantized field operator $\chi$ (or its spatial derivatives), as standard in the holographic correspondence \cite{Harlow:2011ke}.

\subsection{The gluing conditions in JT geometry}

Before moving on to a specific example of \eqref{SvR-JT-average}, we consider the required $C^1$ gluing conditions introduced in Sec. \ref{SvR} for the particular scenario of JT gravity.
This is a fundamental tool to apply in the SvR method which often requires to glue pieces of spacetime of different signature.
The analysis below shows that the quantum gravitational problem as defined on a SK path leads to a well posed variational problem upon providing boundary conditions only on all the asymptotic gravitational boundaries dual to the SK path segments. The semiclassical analysis imposes a set of continuity conditions: field continuity is required off shell to meet basic generating functions properties and conjugated momenta continuity in the complex plane is required on shell. It is necessary to demand $C^1$ continuity for the metric, Dilaton and probe fields over the manifold, i.e. gluing of both the field and its conjugated momenta in the complex plane.

\subsubsection{Deriving the gluing conditions}

Consider the Jackiw-Teitelboim gravity \cite{Teitelboim1983,Jackiw1984} action defined on a generic manifold $M$,
\begin{equation}\label{JT}
16 \pi G_N I_{JT} = \Phi_0 \left( \int_{M}\!\! \sqrt{g} R + 2\int_{M}\!\! \sqrt{h} K \right) + \int_{M}\!\! \sqrt{g} \;\Phi\; (R-2\Lambda) + 2\int_{\partial M}\!\!\sqrt{h} \;\Phi \;(K-1)
\end{equation}
The terms accompanying $\Phi_0$ are topological and stand for the Euler characteristic of the manifold under study, i.e. it is $e^{-\Phi_0(2g+b-2)}$, with $g$ the genus and $b$ the number of boundaries. For most purposes of this work, this term will play no major role. The relevant terms for our analysis are the ones associated with the dynamic Dilaton field $\Phi$. Notice from the second term that the equations of motion for $\Phi$ fix the manifold to be pure AdS$_{2}$, and the only remaining degree of freedom is the Dilaton itself, which from the last term in \eqref{JT} can be seen to represent a reparametrization freedom between the actual physical time in the dual $0+1$ quantum mechanics theory $u$ and the AdS$_2$ time $t$. This ``boundary graviton'' is the only degree of freedom in the theory. One can see that the last term in \eqref{JT} can be written explicitly in terms of the reparametrization field $t(u)$ which have a Schwartzian action  \cite{Sarosi17,Trunin20}. We will explore solutions of the Einstein-Dilaton equations of motion arising from the action above and, in Sec. \ref{sec: wh-sols}, deformations that allow for solutions describing traversable wormholes.

For the purposes of this section, eq. \eqref{JT} should be understood as the manifold $M$ is a single segment of the SK path with defined signature, say Lorentzian. This will contain at least one spacelike boundary $\Sigma$ and one asymptotic boundary $\partial$. Its infinitesimal variation on a manifold with a single timelike $\partial$ and a single spacelike $\Sigma$ boundary\footnote{The existence of $\partial$ and $\Sigma$ boundaries implies that there is a codim-2 boundary in which they meet. This type of terms have been taken into account for SK paths in AdS/CFT in higher $d$ in \cite{SvRL} and in the specific JT context in \cite{Pacman}. We disregard these type of contributions in what follows.} yields,
\begin{align}
16 \pi G_N \delta I_{JT}=& \int \sqrt{g}\left[\frac 12 \;\Phi\; (R-2\Lambda)g^{\mu\nu}-R^{\mu\nu}\Phi+\nabla^\mu\nabla^\nu\Phi - g^{\mu\nu}\nabla^2\Phi\right]\delta g_{\mu\nu} \label{1}\\&
 \quad - \int \sqrt{g} \Phi_0 \left[ R^{\mu\nu}-\frac 12 R g^{\mu\nu} \right] \delta g_{\mu\nu}  + \int \sqrt{g} (R-2\Lambda) \delta \Phi \label{2} \\ &
\qquad + \int_{\partial} \sqrt{h} \left[ 2(K-1)\delta \Phi + \left( n^\nu \nabla_\nu\Phi-\Phi \right)h^{\alpha\beta}\delta h_{\alpha\beta} \right]\label{3}
\\ &
\qquad + \int_{\Sigma} \sqrt{h} \left[ 2 K \delta \Phi +  n^\nu \nabla_\nu\Phi h^{\alpha\beta}\delta h_{\alpha\beta} \right]\label{4}.
\end{align}
An on shell analysis requires $\delta I_{JT}=0$ on the complete SK path. We do this by imposing conditions that make the terms in \eqref{1}, \eqref{2}, \eqref{3} trivial on their own for each segment individually. The terms in the first two lines are bulk terms and are zero on shell by virtue of the equations of motion. The third line correspond are boundary terms but
do not lead to gluing conditions between different pieces of the SK path. For example, fixing the Dilaton and metric on the $\partial$ boundary sets these terms to zero on each segment individually. 

This cannot be done for the terms in \eqref{4} that involve the fields on $\Sigma$, which at best can be put so that they cancel upon a specific gluing between adjacent segments. One could impose some conditions at $\Sigma$ even without introducing any specific (local) action. Basic properties of the generating function force the fields to be continuous on $\Sigma$, i.e. taking two adjacent sample segments $a$ and $b$ from the SK path, one can split and re-glue the partition function as
\begin{equation}
    Z_{SK}=\int d\Phi_\Sigma \;  Z_{a}\Big[\Phi_a|_\Sigma=\Phi_\Sigma\Big]\;  Z_{b}\Big[\Phi_b|_\Sigma=\Phi_\Sigma\Big] \qquad\Rightarrow\qquad \Phi_a|_\Sigma=\Phi_b|_\Sigma \label{GField}
\end{equation}
This conditions is valid off-shell and allows in particular to take $\delta\Phi_a|_\Sigma=\delta\Phi_b|_\Sigma$. As such, two adjacent terms $\delta I_{a,b}$ of the actions above from terms in \eqref{4} combine as
\begin{equation}
\delta I_a + \delta I_b = \int_{\Sigma} \sqrt{h} \left[ 2( K_a-K_b) \delta \Phi +  (n^\nu \nabla_\nu\Phi_a-n^\nu \nabla_\nu\Phi_b) h^{\alpha\beta}\delta h_{\alpha\beta} \right]
\end{equation}
where the relative minus sign comes from the normal vectors being in opposite directions \cite{SvRL}. These contributions can be arranged to cancel themselves by demanding
\begin{equation}
     K_a = K_b\;, \qquad\qquad \Pi_a=\Pi_b \;, \qquad\qquad \text{ on $\Sigma$}\label{GMom}\;,
\end{equation}
where we have used that since $\Sigma$ is a spacelike codim-1 surface, $n^\nu \nabla_\nu\Phi\equiv \Pi$. These conditions are enough to find a candidate saddle to the path integral on the complete SK path upon gluing all segments.

Now, the background solutions we will explore in this work will be time independent on both signatures and of the form,
\begin{equation}\label{samplesol}
ds^2=-h(r)dt^2+h(r)^{-1}dr^2\,, \qquad ds^2=h(r)d\tau^2+h(r)^{-1}dr^2\,, \qquad \Phi=\Phi(r)\,,
\end{equation}
where $h(r)$ is the same function on both metrics and can be taken to be a general function of the radius $r$ for our current purposes. Furthermore, we will always be gluing at $t,\tau$ constant surfaces.
Upon Wick rotation, the induced metric and Dilaton in \eqref{samplesol} remain unchanged at these surfaces so the gluing conditions for the fields \eqref{GField} are met. 
The conditions \eqref{GMom} are also met by the solutions \eqref{samplesol} at $t,\tau$ constant surfaces. The tensor $K_\Sigma$ is
\begin{equation}
\label{CurvExt}
(K_\Sigma)_{\mu\nu}= \frac{1}{2}\mathcal{L}_{n}P_{\mu\nu} = \frac{1}{2} n^{\alpha}\partial_{\alpha}(g_{\mu\nu} - n^{2} n_{\mu}n_{\nu} ) + \partial_{\mu}(n^{\alpha})(g_{\alpha\nu} - n^{2} n_{\alpha}n_{\nu})=0\;,
\end{equation} 
where $n_{\alpha}=\delta_{\alpha,\tau} \sqrt{h(r)}$. As for the Dilaton,
\begin{equation}
    \Pi_\Sigma \equiv n^\nu \nabla_\nu\Phi=\partial_t \Phi=0\;,
\end{equation}
such that we are gluing zero on both sides.

On top of these solutions we will study probe scalar fields $\chi$, which satisfies the Klein-Gordon equations. The gluing conditions for $\chi$ is exactly the same as in the general higher dimensional case so we refer the reader to \cite{SvRC,us1}. We need to $C^1$ glue the field $\chi$ and conjugated momenta $\Pi_\chi$. The continuous gluing will be explicit in our treatment. For future reference we explicitly write the gluing conditions in SK path ordered time,
\begin{equation}\label{Gchi}
    \chi_a=\chi_b\;, \qquad \qquad \Pi_{\chi,a}=\Pi_{\chi,b}\;,  \qquad\qquad \text{ on $\Sigma$}\;.
\end{equation}

\subsection{Correlators in JT}\label{CorrSec3}

To conclude this section we present an example which cover interesting aspects of the SvR construction reduced to the 1+1 JT scenario, eq. \eqref{SvR-JT-average}.  
Interestingly, a naive dimensional reduction of Fig. \ref{Fig:In-Out} is not itself a relevant example for holography: exact AdS$_2$ (with global AdS timelike Killing vector) does not correspond to a sensible Quantum Mechanics with finite energy excitations \cite{Michelson99,Galloway2018}. Accordingly, pure JT gravity does not allow a constant Dilaton profile, which would be the equivalent of a time-like Killing vector in the higher dimensional AdS/CFT examples \cite{Qi}. In this context, we find particularly relevant to provide a fully fledged real time example of JT dual to a single SK path.
Our example can be introduced as a dimensional reduction of the solution presented in \cite{us3,us4} adapted to JT, i.e. including the Dilaton field.

An important comment is due regarding our computation of correlators in JT/SYK correspondence. From the foundational JT/SKY works \cite{Almheiri14,Malda16} it was recognized that the physical correlators in SYK where not directly the ones written in terms of AdS$_2$ time foliation (say $t$) but rather the physical SYK time $u$. Its relation $t(u)$ is defined implicitly by the boundary conditions imposed on the JT problem. More specifically, by the distance at which the AdS$_2$ space is cutoff in each direction. However, it was also found out in these works that since the degree of freedom in this case is a reparametrization mode, for a semiclassical analysis, one can actually follow the general AdS/CFT intuition of extrapolating CFT correlators directly from AdS computations, and then performing a rescaling on the boundary time $t\to t(u)$ \cite{Malda16}. In what follows, we will leave this final step implicit, since it is beyond the point of the physics we want to emphasize, i.e. the real time geometries and correlators allowed by the SvR prescription.

\subsubsection{TFD evolution}
\label{TFDevolution}

\begin{figure}[t]\centering
\begin{subfigure}{0.49\textwidth}\centering
\includegraphics[width=.9\linewidth] {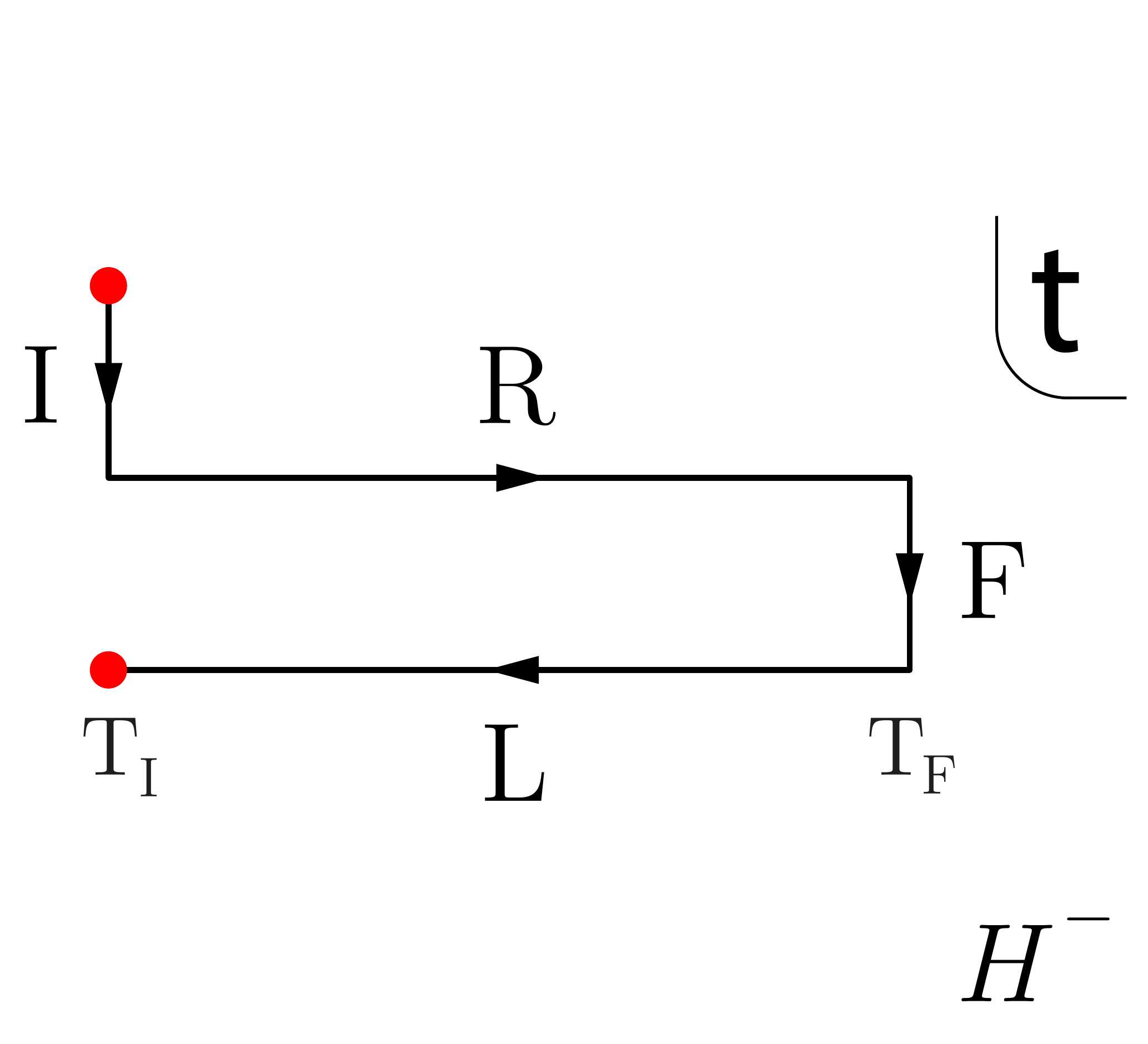}
\caption{}
\end{subfigure}
\begin{subfigure}{0.49\textwidth}\centering
\includegraphics[width=.9\linewidth] {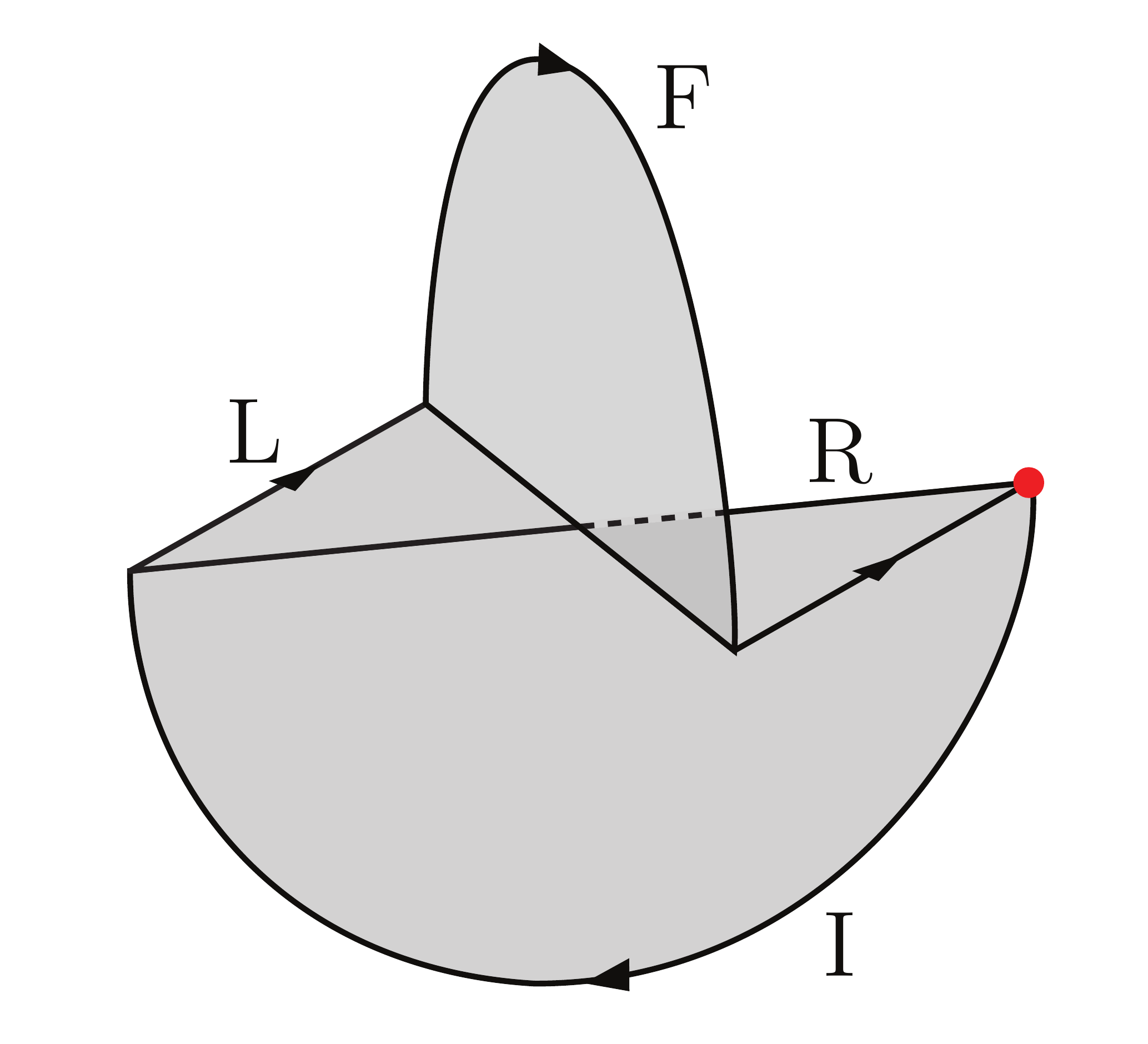}
\caption{}
\end{subfigure}
\caption{(a) SK path associated to a TFD construction in QFT. The total imaginary time evolution is $-i\beta$. The red dots are identified. (b) Bulk saddle point of the TFD SK path. The geometry can be understood as a full Euclidean AdS$_2$ manifold upon a time evolution using a Global Rindler AdS real time evolution at a moment of time reflection symmetry.}
\label{Fig:TFD}
\end{figure}

Here we present a time TFD-like evolution of a thermal circle of length $\beta$. The SK path associated to this problem can be seen in Fig.\ref{Fig:TFD}(a) and previous holographic work on this path for standard AdS/CFT can be found in \cite{HerzogSon,us3,us4}, see also \cite{Umezawa} for a QFT introduction. The path extends forwards in real time, evolves $-i\beta/2$ in imaginary time and then comes backwards to the initial time before closing the thermal circle with a final $-i\beta/2$ evolution. 

The dual geometry is built from two Lorentzian AdS$_2$ Schwarzschild BH exteriors (or equivalently in AdS$_2$, two Rindler patches) dubbed $R,L$ and an Euclidean BH manifold halved in two pieces $I,F$ as shown in Fig.\ref{Fig:TFD}(b). 
The metrics are
\begin{equation}\label{metric}
ds^2=-(\rho^2-1)d \sigma^2+\frac{d\rho^2}{(\rho^2-1)} \;,
\qquad\qquad ds^2 = (\rho^2-1) d\varsigma^2 + \frac{d\rho^2}{(\rho^2-1)}\;,\qquad\qquad \rho\in[1,\infty)
\end{equation}
for the Lorentzian and Euclidean regions respectively. The Lorentzian time is taken $T_-\leq \sigma\leq T_+$ for both pieces, a positive time evolution in $R$ and $L$ running in opposite directions, see Fig. \ref{Fol}, as mandated by the standard TFD construction.
This is equivalent to stating that $H^-=H_R-H_L$ is taken as the global Hamiltonian for the system\footnote{For the same SK path, the $H^+=H_R+H_L$ choice is difficult to approach in the JT/SYK correspondence. Essentially, the JT EOMs do not support a solution invariant wrt the AdS$_2$ Global time Killing vector ( there is no constant Dilaton solution) such that the Euclidean pieces that close the path at $T_{I/F}$ would depend explicitly on $\Delta T$. Finding these explicit solutions goes beyond the scope of this work. This is reminiscent of the Global time evolution of the TFD state prepared by the gravity Euclidean path integral for a Black Hole, where no Global $H^+$ time Killing vector is present, see e.g. \cite{eternal}. \label{footHplus}},
see \cite{Umezawa,us3,us4} for more details.
The regions $I,F$ have the range $-\pi \leq \varsigma \leq 0$ and $0 \leq \varsigma \leq \pi$ respectively. 
The two Euclidean halves are cut open at constant $\varsigma$-curves and pieces are glued to the constant $\sigma$-curves, thus meeting the continuity conditions \eqref{GField} and \eqref{GMom}. In this coordinates the solution for the Dilaton is
\begin{equation}
\Phi=\rho\;\phi_b,
\end{equation}
where $\phi_b$ is a constant boundary value. The solution is time independent, thus also meeting the continuity conditions. This completes the description of the background gravity solution.

We now probe our construction with a real scalar field $\chi(\sigma, \rho)$ with action, equation of motion and boundary condition given by
\begin{equation}\label{R-EDM}
    I_{KG} = -\int \sqrt{g}(\partial_\mu \chi \partial^\mu \chi+m^2 \chi^2)\:,\qquad\qquad \left(\square-m^2\right)\chi=0, \qquad\qquad \chi(\rho,\sigma)|_{\partial}\sim \rho^{\Delta-1}\chi_R(\sigma)
\end{equation}
respectively. The conformal dimension of the boundary operator $\Delta$ is defined through the standard relation $m^2=\Delta(\Delta-1)$.
By expanding the solution in Fourier modes of frequency $\omega$ one gets the radial equation,
\begin{equation} 
  \left((\rho^2-1)\chi'(\rho)\right)'  = \left( \Delta(\Delta-1) -\frac{\omega^2}{\rho^2-1} \right)\chi(\rho)\;.
\end{equation}
A solution meeting the boundary conditions (NN modes) can be written as,
\begin{equation}\label{R-sol}
    \chi_{NN}(\sigma,\rho) = \frac{1}{2\pi}\int_{-\infty}^{\infty} d \sigma' \int_{-\infty}^{\infty} d\omega \; e^{-i\omega (\sigma-\sigma')} \left( R^+_\omega \; p^{+}_{\Delta\omega}(\rho) + R^-_\omega \; p^{-}_{\Delta\omega}(\rho) \right) \chi_R(\sigma'),\qquad\qquad R^+_\omega+R^-_\omega=1
\end{equation} 
with the eigenfunctions $p^{\pm}_{\Delta\omega}$ defined as
\begin{align}\nn
p^{\pm}_{\Delta\omega}(\rho)&\equiv 2^{\Delta -1}
e^{\pm \frac{\pi  \omega }{2}} 
\frac{ \Gamma (\Delta ) \Gamma (\Delta \pm i \omega )}{\Gamma (2 \Delta -1)} P_{\Delta-1}^{\mp i\omega}(\rho)
\simeq 1\times \rho^{\Delta-1} + \dots + \frac{2^{1-2 \Delta } \Gamma \left(\frac{1}{2}-\Delta
   \right) \Gamma (\Delta \pm i \omega )}{\Gamma
   \left(\Delta -\frac{1}{2}\right) \Gamma (-\Delta \pm i \omega +1)}  \times \rho^{\Delta}+\dots
\end{align}
where $P_{n}^{m}(\rho)$ are the associated Legendre polynomial and the custom $p^{\pm}_{\Delta\omega}$ notation stands for the function containing its poles only in the upper/lower half $\omega$ plane at $\omega_n=\pm i (\Delta +n)$. The $R_\omega^\pm$ notation is convenient for our purposes and one can see that $R_\omega^+/R_\omega^-$ can
be interpreted as the relative weight of outgoing and infalling modes through the horizon \cite{us3}. 
Notice that so far there are no restrictions on the $R_\omega^+/R_\omega^-$ quotient, which means that \eqref{R-sol} is not yet uniquely defined as a solution. Fixing this quotient amounts to choosing a particular time ordering for the real time correlator \cite{HerzogSon,SvRL,us3}.

Furthermore, a full set\footnote{The physical normalizable mode basis is actually not continuous. A discrete set of N modes rises once the inner product between functions is correctly orthonormalized, see e.g. \cite{Kenmoku}.} of normalizable (N) modes of the form 
\begin{equation}\label{R-solN}
    \chi_N(\sigma,\rho)=\frac{1}{2\pi}\int_{-\infty}^{\infty} d\omega \; e^{-i\omega \sigma} N_\omega\left(  p^{+}_{\Delta\omega}(\rho) - p^{-}_{\Delta\omega}(\rho) \right)
\end{equation} 
with arbitrary $N_\omega$ can still be added to the solution. 

Analogous general solutions with undetermined coefficients can be found for each of the $I,R,F,L$ regions in terms of their own asymptotic sources. This freedom in the solutions is completely determined once the solutions in the different piece-wise bulk duals are glued together according to \eqref{Gchi}. 

As an example, consider a problem in which only asymptotic boundary conditions on $R$ are turned on and asymptotic boundary conditions are turned off in $I,L,F$. For region $R$ the solution $\chi_R$ should be a sum of \eqref{R-sol} and \eqref{R-solN}. On the other hand, for the $I,L,F$ regions the solution should contain only N modes similar to \eqref{R-solN},
\begin{align}
    \chi_I(\varsigma,\rho)&=\frac{1}{2\pi}\int_{-\infty}^{\infty} d\omega \; e^{-\omega \varsigma} I_\omega\left(  p^{+}_{\Delta\omega}(\rho) - p^{-}_{\Delta\omega}(\rho) \right) \nn\\
    \chi_L(\sigma,\rho)&=\frac{1}{2\pi}\int_{-\infty}^{\infty} d\omega \; e^{-i\omega \sigma} L_\omega\left(  p^{+}_{\Delta\omega}(\rho) - p^{-}_{\Delta\omega}(\rho) \right) \label{NmodeExample}\\
    \chi_F(\varsigma,\rho)&=\frac{1}{2\pi}\int_{-\infty}^{\infty} d\omega \; e^{-\omega \varsigma} F_\omega\left(  p^{+}_{\Delta\omega}(\rho) - p^{-}_{\Delta\omega}(\rho) \right)\nn
\end{align}
when no sources are present. Imposing $C^1$ field continuity \eqref{Gchi}
in this foliation imposes \cite{us3},
\begin{eqnarray}\label{TFD-bc}
\chi_R=\chi_I\,,  & \quad
-i\partial_t\chi_R=\partial_\varsigma\chi_I\,, 
& \qquad\text{on \quad$\sigma=T_I$,\quad $\varsigma=0$}\nn \\
\chi_R=\chi_F\,, & \quad
-i\partial_t\chi_R=\partial_\varsigma\chi_F\,, 
& \qquad\text{on \quad$\sigma=T_F$,\quad $\varsigma=0$} \nn\\
\chi_L=\chi_I\,,& \quad
-i\partial_t\chi_L=\partial_\varsigma\chi_I\,, 
&\qquad\text{on \quad$\sigma=T_I$,\quad $\varsigma=-\pi$} \nn\\
\chi_L=\chi_F\,,& \quad
-i\partial_t\chi_L=\partial_\varsigma\chi_F\,, 
& \qquad\text{on \quad$\sigma=T_F$,\quad $\varsigma=\pi$}\;.
\end{eqnarray}
One finds for $R$
\begin{equation}\label{Lpm}
    R^+_\omega=\frac{-1}{e^{2\pi\omega}-1} \qquad\qquad  R^-_\omega=\frac{e^{2\pi\omega}}{e^{2\pi\omega}-1}\qquad\qquad N_\omega=0
\end{equation}
and for the other regions
\begin{equation}
    I_\omega=-e^{-i \omega T_I} \tilde\phi_{R;\omega}\frac{1}{e^{2\pi\omega}-1} \qquad L_\omega=-\tilde\phi_{R;\omega}\frac{e^{2\pi\omega}}{e^{2\pi\omega}-1}\qquad F_\omega=-e^{-i \omega T_F} \tilde\phi_{R;\omega}\frac{e^{2\pi\omega}}{e^{2\pi\omega}-1}
\end{equation}
with $\tilde\phi_{R;\omega}$ the Fourier Transform of the $\phi_R(\sigma)$.
We stress that all coefficients are completely determined after gluing. This fixes both the initial state of our theory (the TFD state in this case) and the correlator time ordering (Feynman ordering) as we check below. Notice that crucially $I_\omega$ and $F_\omega$ are exponentially suppressed in $\omega\to\pm\infty$ such that the $\omega$ integral in the anzats can be seen to converge. 

Correlators can be computed using the prescribed eq. \eqref{SvR-JT-average} with on shell action eq. \eqref{R-EDM} after standard holographic renormalization, i.e.
\begin{align}\nn
    -i S^0&=\frac{i}{2} \int_\partial \sqrt{\gamma} \chi (n^\mu \partial_\mu \chi)\\ 
    &=\int d\sigma d\sigma' \chi_R(\sigma) \left[\frac{ 4^{-\Delta } \Gamma \left(\frac{3}{2}-\Delta
   \right)}{i \pi  \Gamma \left(\Delta -\frac{1}{2}\right)}\int_{-\infty}^{\infty} d\omega \; \frac{e^{-i\omega (\sigma-\sigma')}}{e^{2\pi\omega}-1} \times \right.\nn\\
   &\qquad\qquad\qquad\qquad\qquad\qquad\qquad\qquad\times\left. \left( - \; \frac{ \Gamma (\Delta + i \omega )}{ \Gamma (1-\Delta + i \omega)} + e^{2\pi\omega}\; \frac{ \Gamma (\Delta - i \omega )}{ \Gamma (1-\Delta - i \omega )} \right)\right] \chi_R(\sigma')\;,\label{onchell}
\end{align}
where the (Feynman ordered) correlator is captured in the squared brackets above. We will proceed with the computation in two steps. 

The first step is by noticing that the $\omega$ integral can be explicitly carried out via Residue Theorem, closing the complex path from above or below depending on the sign of $(\sigma-\sigma')$. The singularity structure of the integrand contains two families of poles: (i) at $\omega=\pm i (\Delta+n)$ coming from the $\Gamma(\Delta\pm i\omega)$ functions and (ii) at $\omega=\pm in$ coming from the $e^{2\pi\omega}-1$ denominator. We now show that the second set of poles never contribute to the correlator regardless of the sign of $(\sigma-\sigma')$. 
To see this, consider closing the integral from below ($(\sigma-\sigma')>0$) and take the $\omega=-in$, $n\in\mathbb{N}$. The residue of $(e^{2\pi\omega}-1)^{-1}$ is 1 for all $n$, so their contribution can be isolated as
\begin{equation}
    \sum_{n\geq0}e^{n(\sigma-\sigma')}\left(  - \; \frac{ \Gamma (\Delta +n )}{ \Gamma (1-\Delta +n)} + \frac{ \Gamma (\Delta - n )}{ \Gamma (1-\Delta -n )} \right)= 
    \sum_{n\geq0}\frac{\pi\; e^{n(\sigma-\sigma')}(-1)^{n}}{\Gamma (1-\Delta -n)\Gamma (1-\Delta +n)\sin(\pi\Delta)}
    \left( - 1+1 \right)=0
\end{equation}
which is identically zero term by term. A similar cancellation on the set of ``thermal'' poles was seen in \cite{SvRL,us3} and more recently in \cite{Loganayagam22}.
As for the (i) $\omega= -i (\Delta+n)$ poles, for $(\sigma-\sigma')>0$ we only pick the second term in brackets in \eqref{onchell} to obtain,
\begin{align}\label{pre-TFD-RR}
\langle {\cal O}_R(\sigma){\cal O}_R(\sigma')\rangle &= \frac{2  \Gamma \left(\frac{3}{2}-\Delta
   \right)}{4^{\Delta }i \pi  \Gamma \left(\Delta -\frac{1}{2}\right)} \int_{-\infty}^{\infty} d\omega \; \frac{e^{-i\omega (\sigma-\sigma')}}{e^{2\pi\omega}-1} \left( - \; \frac{ \Gamma (\Delta + i \omega )}{ \Gamma (1-\Delta + i \omega)} + e^{2\pi\omega}\; \frac{ \Gamma (\Delta - i \omega )}{ \Gamma (1-\Delta - i \omega )} \right)\\ \nn
&= \frac{2  \Gamma \left(\frac{3}{2}-\Delta
   \right)}{4^{\Delta }i \pi  \Gamma \left(\Delta -\frac{1}{2}\right)} \int_{-\infty}^{\infty} d\omega \; e^{-i\omega (\sigma-\sigma')}\;\frac{e^{2\pi\omega}}{e^{2\pi\omega}-1} \; \frac{ \Gamma (\Delta - i \omega )}{ \Gamma (1-\Delta - i \omega )}\;,\qquad\qquad (\sigma-\sigma')>0 \\ \nn
&= \frac{2  \Gamma \left(\frac{3}{2}-\Delta
   \right)}{4^{\Delta }i \pi  \Gamma \left(\Delta -\frac{1}{2}\right)}  \frac{2 \pi i }{1-e^{2 i \pi  \Delta }}\sum_{n=0}^{\infty} \frac{(-1)^n }{  n! }\frac{e^{-(n+\Delta
   ) \left(\sigma-\sigma'\right)}}{\Gamma (1-n-2 \Delta )}\\\nn
&= \frac{\left(2\Delta -1\right) \Gamma (\Delta )}{\sqrt{\pi } \Gamma \left(\Delta -\frac{1}{2}\right)}\; \frac{1}{\left[\cosh\left(\sigma-\sigma'\right)-1\right]^{\Delta}}
\end{align}
The $(\sigma-\sigma')<0$ scenario follows similarly (one rather picks the first term in brackets in \eqref{onchell} in closing the integral from above) and yields the same result.

The second step is to determine the causal propagation the SK path is prescribing. Notice that we have already shown that the integral is non trivial whenever $|\sigma-\sigma'|\neq0$, so our correlator is neither retarded nor advanced.
One can see that the results we have obtained so far are equivalent to consider an exponential in time regulated as $(\sigma-\sigma')(1-i\epsilon)$, $\epsilon>0$, which corresponds to standard Feynman ordering. One can confirm this claim by going back to \eqref{onchell} and check that for $(\sigma-\sigma')>0$ one can still drop the first term going to the second line. Then, the factors $\exp\{-i \omega (\sigma-\sigma')(1-i\epsilon)\}\sim \exp\{ - \omega \epsilon \}$ 
correctly regulate the $\omega^{-(2\Delta-1)}$ divergence as $ \omega \to + \infty $ in the integrand. For $(\sigma-\sigma')<0$ the Feynman regulator behaves as $e^{+\omega\epsilon}$, correctly regulating the integrand oscillations as $\omega \to -\infty$.
Finally we get
\begin{align}
\langle {\cal O}_R(\sigma){\cal O}_R(\sigma')\rangle 
&=  \frac{\left(2\Delta -1\right) \Gamma (\Delta )}{\sqrt{\pi } \Gamma \left(\Delta -\frac{1}{2}\right)}\;  \frac{1}{\left[\cosh\left((\sigma-\sigma')(1-i\epsilon)\right)-1\right]^{\Delta}}\label{TFD-RR}
\end{align}

Similar computations can be done by turning on asymptotic sources for the other regions $I,L,R$. The linearity of the problem allows to solve for only one source turned on at a time, and then find the general solution by addition. As reviewed in Sec. \ref{Setup}, boundary sources on the Lorentzian regions are usually put to zero after derivation and are thought as a tool to compute correlators. On the other hand, Euclidean sources can be left turned on to consider holographic excited states in the geometry and probe their real time evolution. 
We will not carry all the computations since most of the important pieces stem from the example above. We thus only present the results for correlators and expectation values of the available observables, i.e. ${\cal O}_R,{\cal O}_L$, in this simple model.

The correlators are
\begin{align*}
\langle {\cal O}_L(\sigma){\cal O}_L(\sigma')\rangle =\langle {\cal O}_R(\sigma){\cal O}_R(\sigma')\rangle^*\;, 
\qquad\langle {\cal O}_L(\sigma){\cal O}_R(\sigma')\rangle = 
\frac{\left(2\Delta -1\right) \Gamma (\Delta )}{\sqrt{\pi } \Gamma \left(\Delta -\frac{1}{2}\right)}\;  \frac{1}{\left[\cosh\left(\sigma-\sigma'\right)+1\right]^{\Delta}}
\end{align*}
where the resulting time ordering in the $\langle {\cal O}_L(\sigma){\cal O}_L(\sigma')\rangle$ correlator results reverse-time ordered. Notice that correlators with operators inserted on opposite sides does not have any singularities, signaling the lack of causal communication between boundaries. 

Expectation values of the operators yield via BDHM prescription \eqref{BDHM2-ave},
\begin{align}\label{TFD-R}
  \langle \chi_F| {\cal O}_R(\sigma) |\chi_I \rangle  &= \frac{ 2 \Gamma \left(\frac{3}{2}-\Delta \right)} {4^{\Delta }  i \pi  \Gamma \left(\Delta -\frac{1}{2}\right)} 
   \int_{-\infty}^{\infty} d\omega \; \frac{e^{-i\omega \sigma}}{e^{2\pi\omega}-1}  \frac{ \Gamma (\Delta + i \omega )}{ \Gamma (1-\Delta + i \omega)} \times\\
   & \qquad\qquad\qquad\qquad\qquad\qquad\times\left( -\int_{-\pi}^0 d\varsigma \chi_I(\varsigma) e^{-\omega \varsigma} + \int_0^{\pi} d\varsigma \;e^{2\pi\omega} \chi_F(\varsigma) e^{-\omega \varsigma} \right) \;\nn\\ 
   &= 
   \frac{\left(2\Delta -1\right) \Gamma (\Delta )}{\sqrt{\pi } \Gamma \left(\Delta -\frac{1}{2}\right)}\; 
   \left( \int_{-\pi}^0 d\varsigma\frac{\chi_I(\varsigma)}{\left[\cosh\left(\sigma-i\varsigma\right)-1\right]^{\Delta}} + \int_{0}^\pi d\varsigma \frac{\chi_F(\varsigma)}{\left[\cosh\left(\sigma-i\varsigma\right)-1\right]^{\Delta}}\right)
\end{align}
\begin{equation}
  \langle \chi_F| {\cal O}_L(\sigma) |\chi_I \rangle  =  
  \frac{\left(2\Delta -1\right) \Gamma (\Delta )}{\sqrt{\pi } \Gamma \left(\Delta -\frac{1}{2}\right)}\; \left( \int_{-\pi}^0 d\varsigma\frac{\chi_I(\varsigma)}{\left[\cosh\left(\sigma+i\varsigma\right)-1\right]^{\Delta}}+\int_{0}^\pi d\varsigma \frac{\chi_F(\varsigma)}{\left[\cosh\left(\sigma+i\varsigma\right)-1\right]^{\Delta}}\right)
\end{equation}
which can be computed for any profile of Euclidean sources. The first expression in \eqref{TFD-R} should be contrasted with \eqref{BDHM2-ave} shows that one can decompose the excitation in the N modes of the geometry and follow them individually. 

We now make two comments on the results of the example above:

\paragraph{$\bullet$ Only the trivial solution in absence of sources as a check of the gluing conditions:}

A generic SK path split in several real and imaginary time pieces leads to many bulk region pieces which should be $C^1$ glued to each other. 
We propose a consistency check for these gluing conditions based on the fact that (at least near the boundary, since the bulk interior might develop a non-trivial topology) one is implementing a $C^1$ gluing on a closed complex-time path of known period $-i\beta$.

The trick is to study the problem of finding a bulk solution when all asymptotic boundary sources are turned off. This amounts to considering a solution in $R$ as \eqref{R-solN} and solutions eq. \eqref{NmodeExample} for the other regions.  
The only freedom in these solutions are its coefficients $N_\omega,I_\omega,L_\omega,F_\omega$ that are fixed by eqs. \eqref{TFD-bc}.

Since the SK problem is well posed, its solution being unique upon giving boundary conditions on the asymptotic boundary, and we have turned off all sources, it stems that the anzats $\chi=0$ must be the only consistent solution. We can use this condition to check our derived gluing conditions. If correct, the gluing of all N modes should impose
\begin{equation}
    N_{\omega} = e^{-\beta \omega} N_{\omega}
\end{equation}
and the same condition repeated for all other coefficients. Its solutions are either $-i \beta\, \omega \sim 2 \pi n$ or $N_{\omega}=0$. The former solution, i.e. an expansion on the Matsubara frequencies of the pure Euclidean scenario, is crucially inconsistent in real time. The reason for it is that pure imaginary $\omega$ in the real time scenario forces the asymptotic behaviour of the solution to change such that it no longer decays at the conformal boundary, i.e. the N modes become divergent at the boundary. More precisely, $p^{\pm}_{\Delta\omega}(\rho)\sim e^{\pm\omega\ln \rho}$ at $\rho\to\infty$ such that for imaginary frequencies a cancellation between these, as we have built for the real time scenario is no longer possible. This inconsistency also appears for any dimension $d>1$, see e.g. \cite{us3,us4}. 
We thus conclude that the only consistent solution is $N_\omega=I_\omega=L_\omega=F_\omega=0$, i.e. there are no pure N modes in the geometry. We conclude that the real time pieces of the problem impose non-trivial conditions that can help to check if we have written our gluing conditions correctly in our preferred time foliation.

\paragraph{$\bullet$ Other correlator orderings:}

The in-depth analysis on the location of the complex poles in the field solution we carried in this example allows to go beyond the Feynman propagator. Specifically, by coming back at eq. \eqref{pre-TFD-RR} one can extract both the retarded and advanced correlation functions
\begin{align*}
Ret\{\langle {\cal O}_R(\sigma){\cal O}_R(\sigma')\rangle\} &= \frac{ 2 \Gamma \left(\frac{3}{2}-\Delta \right)} {4^{\Delta }  i \pi  \Gamma \left(\Delta -\frac{1}{2}\right)}  \int_{-\infty}^{\infty} d\omega \; e^{-i\omega (\sigma-\sigma')} \frac{e^{2\pi\omega}}{e^{2\pi\omega}-1} \; \frac{ \Gamma (\Delta - i \omega )}{ \Gamma (1-\Delta - i \omega )} \\
   &=\Theta(\sigma-\sigma')\times\frac{\left(2\Delta -1\right) \Gamma (\Delta )}{\sqrt{\pi } \Gamma \left(\Delta -\frac{1}{2}\right)}\;  \frac{1}{\left[\cosh\left(\sigma-\sigma'\right)-1\right]^{\Delta}} 
\end{align*}
\begin{align*}
Adv\{\langle {\cal O}_R(\sigma){\cal O}_R(\sigma')\rangle\} &= \frac{ 2 \Gamma \left(\frac{3}{2}-\Delta \right)} {4^{\Delta }  i \pi  \Gamma \left(\Delta -\frac{1}{2}\right)} \int_{-\infty}^{\infty} d\omega \; e^{-i\omega (\sigma-\sigma')} \frac{(-1)}{e^{2\pi\omega}-1}  \; \frac{ \Gamma (\Delta + i \omega )}{ \Gamma (1-\Delta + i \omega)} \\
   &= \Theta(\sigma'-\sigma)\times\frac{\left(2\Delta -1\right) \Gamma (\Delta )}{\sqrt{\pi } \Gamma \left(\Delta -\frac{1}{2}\right)}\;  \frac{1}{\left[\cosh\left(\sigma-\sigma'\right)-1\right]^{\Delta}}
\end{align*}
which can be checked to be correct by noticing that both provide the correct expression for the correlator poles only on the upper/lower half $\sigma$-plane, i.e. providing the correct Heaviside $\Theta$ functions.

\section{Wormholes in holographic 2d gravity}
\label{sec: wh-sols}

In this Section we are going to study many aspects of two dimensional gravity, as being dual to suitable quantum systems defined on disconnected boundaries. Starting from the fact already observed in the previous Section, that any classical 2d spacetime $M$ (with arbitrary piecewise signature) have $b$ closed curves as boundaries $ \partial M = {\cal C}^1 \sqcup \dots {\cal C}^1$, we construct and study the real-time holographic prescriptions for $b>1$, involving (or not) averaging on coupling constants.
In doing this, one must deal with the so-called factorization problem and emphasize its implications on that holographic prescriptions and dual quantum theories. 

The natural generalization of the proposal JT/SYK is to consider the tensor product of $b$ copies of the SYK models, but in order to get interesting wormhole spacetimes consistent with holography, we must consider suitable generalizations of JT model as dual gravity. 

\subsection{On the factorization problem and its resolution}

The so-called factorization problem can be summarized as follows. For simplicity, consider a manifold with two boundaries $b=2$, the prescription \eqref{SvR-JT} reads,
\begin{equation}\label{factorization}
\text{Tr} \, \;U [{\cal C}_1] \;U [{\cal C}_2] = Z_{grav} (\kappa_1 , \kappa_2) \qquad\qquad \partial M \equiv {\cal C}_1\sqcup{\cal C}_2\qquad \kappa_{1,2} = \chi|_{{\cal C}_{1,2}}
\end{equation} 
Since the operators $U [{\cal C}_1]$, $U [{\cal C}_2]$ act on the Hilbert spaces ${\cal H}_1$, ${\cal H}_2$ respectively, the left hand side of this equation \emph{factorizes} as
$$\text{Tr} \, \;U [{\cal C}_1] \;U [{\cal C}_2] =
(\text{Tr}_1 \,  U [{\cal C}_1])\;(\text{Tr}_2 \,  \;U [{\cal C}_2]\, ) \;,$$
while inconsistently, the rhs involves \emph{a sum} over different spacetime topologies $M_1 \sqcup M_2 $, plus all the connected manifolds with two boundaries and genus $g\geq1$, which \emph{does not factorize}. It is important emphasize here that eq. \eqref{factorization} stands for the exact expression, valid to all quantum level/order, and the gravitational path integral must not be understood in terms of any semiclassical or saddle point approximation. 

This is nothing but a more refined form of the argument proposed in \cite{ABS}\footnote{In that old version of the argument it was shown that, in certain wormhole AdS$_5$-spacetimes, the CFT state ($\sim U^{\otimes 2}$) should be described by a 
thermal density matrix $e^{-\beta H}$, where $H$ should contain a term \emph{coupling} both boundaries. }
to conclude that the assumption $U [{\cal C}_1]$, $U [{\cal C}_2]$ acting on the Hilbert spaces ${\cal H}_1$, ${\cal H}_2$ separately should be incorrect, and therefore, at least a part of the total Hamiltonian ${\cal H}_{1}$ should involve operators acting on $\;{\cal H}_{2}$ and vice-versa.
It is equivalent to the presence of (coupling) terms involving operators of the two field theories in the Hamiltonian, e.g. $\propto {\cal{O}}_1 {\cal{O}}_2 $. Later, this argument was drastically enforced in \cite{Gao}, where it was shown that these type of (double trace) terms in the boundary Hamiltonian slightly deform the dual geometry to allow traversable wormholes.

In more recent years, it has been argued that the factorization problem would be absent in the context of averaging theories as SYK in $0+1d$, where the boundary field theory, supposed to be dual to JT gravity (or at least certain effective dof´s), is obtained as a suitable average on certain family of randomly coupled theories \cite{Saad:2021uzi}. In what follows we are going to formulate more precisely the prescriptions for this type of holographic duality, and will try to explain how the averaging mechanism can avoid the factorization problem, in fact, we will show that theories ${\cal H}_1$, ${\cal H}_2$ with random parameters on each boundary give rise to  an effective coupling between them. 

In summary, there are at least two ways to solve this apparent conflict, namely: \textbf{a)} in a purely holographic theories (without any averaging) one must accept the presence of terms in the theory that couple the fields on disconnected components of the asymptotic boundary \cite{ABS, Gao}; otherwise, \textbf{b)} one should relax the standard holographic dictionary by assuming some proper type of averaging on the lhs of the formula \eqref{SvR-JT}.
However, we are going to finally show that in the SYK case the second option yields to the first one in an effective sense.

\subsubsection{SYK$^b$/JT: the real-time prescription for $b$ disconnected boundaries}
\label{Sec:<SvR-manyb>}

We have argued in Sec. \ref{Sec:<SvR>} that any 2d-spacetime $M$ with $b$ boundaries, must be holographically described by $b$ closed curves where the (averaged) evolution operators are defined on. Thus, the prescription \eqref{SvR-JT-average} can be generalized by doing the following natural replacement
\begin{equation}\label{SYK-average-manyb}
 \text{Tr} \Bigg\{ \bigotimes_{A=1}^b \;U [J_A, \kappa_A, {\cal C}_A]\Bigg\}
 ~ \to  ~ \langle\text{Tr} \, U [\kappa, {\cal C}]\rangle \equiv \int dJ_1 \dots dJ_b \, G(J_1,\dots, J_b)\, \text{Tr} \Bigg\{ \bigotimes_{A=1}^b \;U [J_A, \kappa_A, {\cal C}_A]\Bigg\}
\end{equation}
On the the lhs the tensor product is because we are initially assuming that the operators $U [{\cal C}_A]$ act on their own Hilbert spaces ${\cal H}_A$ respectively, $A=1,\dots,b$.
So the novel modified holographic formula, that substitutes the SvR recipe capturing the field theory averaging, $\langle$ SvR $\rangle$, expresses as follows
\begin{equation}\label{SvR-JT-average-manyb}
 \langle\,\text{Tr}\, U\left[\kappa_1, \dots, \kappa_b\, ; \,{\cal C}_1\sqcup \dots \sqcup {\cal C}_b \right] \rangle = Z_{JT} (\kappa_1, \dots, \kappa_b)
\end{equation}
The left hand side stands for an average on $\text{Tr}\, U$'s which is \emph{equivalent} to a path integral on a closed curve. In forthcoming Sections, this prescription is going to be implemented using some specific 2d-wormhole solutions already studied in the literature.

Let us remark here a couple of important constraints on the distribution $G_b \equiv G(J_1,\dots, J_b)\,$ that appears in this formula: notice first that in the large $N$ limit, the right hand side is given by the saddle point approximation, thus, if the bulk theory is strictly JT gravity, whose fields are Dilaton and metrics with no additional back-reacting fields, the right hand side factorizes as
\be \label{saddleJT} Z_{JT} (\kappa_1, \dots, \kappa_b) \approx e^{iI_{JT} (\kappa_1, {\cal C}_1)}\dots e^{iI_{JT} (\kappa_b, {\cal C}_b)} ,\ee 
and each of these theories is dual to a single SYK model \eqref{SvR-averaged}, 
which implies that to large $N$ the leading contribution to $G_b$ must be a product of $b$ normal distributions as \eqref{gaussian}.
On the other hand, one can be tempted to take this as the more natural distribution even at quantum level; nevertheless, if the field theories on each disconnected boundary are decoupled among them, the lhs of \eqref{SYK-average-manyb} would factorize, and again, it would contradict the quantum (JT) gravity path integral on the right hand side. Therefore, in order to avoid this paradox we conclude that in SYK$^b$/JT duality, the distribution \emph{only} can factorize in the large $N$ limit.

These two facts can be summarized in the following expression,
\be \label{GpureJT} 
G(J_1 ,\dots J_b) \, \to\,  G(J_1)  \dots G(J_b) \qquad \text{as} \qquad N\gg1 \; ,
\ee
where each factor on the right is given by \eqref{gaussian}, while other (sub-leading in $1/N$) contributions cannot factorize\footnote{However, we will see \eqref{GpureJT}
that although $G_b$ can be written exactly as a product of distributions, the factorization issue can be avoided by introducing more random couplings in the SYK model}. 
Notice, however, that there are more general frameworks where the two dimensional gravity admit connected wormholes as the dominant classical solution. In these scenarios the first part of this argument fails because the saddle point approximation \eqref{saddleJT} does not factorize. In this case the leading contribution (to large N) must be very different from \eqref{GpureJT}. We discuss some realizations of ensembles that violate \eqref{GpureJT} below.

\subsubsection{Random SYK$^b$ models and the rigidity constraint}
\label{Sec:Rigidity-Wh}

If we consider that the quantum (random) systems living on different (disconnected) boundaries are similar, it is quite natural that the general properties and constants that characterize each one are the same. Thus, the function $G(J_1,\dots, J_b)$, as well as the actions, should factorize in $b$ equal random models characterized by $\sigma_1, \dots, \sigma_b \equiv \sigma$ and vanishing mean values. However as we just argued, this runs into trouble with the averaging-resolution of the factorization puzzle. 

In that sense, one can assume some specific form of the function $G_b$, implementing some constraint between the random couplings $J$s defined on the different boundaries. In fact, they can be related in some simple way, e.g, through a set of linear relations
$$ \sum_{B=1}^b\; c_{AB}\, J_B + c_A =0 \;\;\qquad A, B=1,\dots b \;.$$ which we can define as \emph{rigidity relations}. This is a further requirement,
whose origin from fundamental features of the SYK$^b$ quantum system is not within the scope of this article.
The simplest form of this constraint is 
\be\label{rigidity} J_1 =\dots =J_b \,.\ee
These properties can be summarized and described in the generalized $\langle$SvR$\rangle$ recipe \eqref{SvR-JT-average-manyb}, 
by defining
\be\label{rigid-G-b-bdys} G(J_1,\dots, J_b )\equiv\;\; \prod^{b-1}_{A=1}\, \int_{-\infty}^\infty \, dJ_A \; G_{\sigma}(J_A) \;\delta(J_A - J_{A+1}) \; \ee 
which from now on, will be simply referred to as \emph{random rigidity}. We are going to see below and in the forthcoming toy models that this type of distributions induce effective couplings between the boundaries, which would be consistent with wormhole-like gravitational saddle points. Note that this distribution does not satisfy \eqref{GpureJT}. 

\paragraph{$\bullet$ Rigid random models and dual wormhole geometry:}

Consider the simplest case $b=2$. The specific SYK model on each boundary is described by the action 
\be\label{SYK-A}
I[J_A, \Psi_A] \equiv \int_{{\cal C}_A} d\theta\; \left( \frac{1}{2}\psi_A\;  \cdot \; \dot \psi_A + \frac{1}{4!}\sum_{i,j,k,l=1}^N \,J_{A ijkl} \psi_A^i \psi_A^j\psi_A^k\psi_A^l \right)\;;\qquad\qquad A=1,2\;.
\ee 
on a closed complex-path ${\cal C}_A$. The dot of the kinetic term denotes sum over repeated  $i =1, \dots, N $, (i.e $( x_i \cdot y_i )= \sum_{i=1}^N x_i y_i$).
The lhs of the recipe \eqref{SvR-JT-average-manyb} in this case reads
\begin{align}\label{SYKb2-rigidaverage}
 \langle\,\text{Tr}\, U\left[\kappa_1,  \kappa_2\, ; \,{\cal C}_1\sqcup {\cal C}_2 \right] \rangle &=
\int \prod_A {\cal D} \psi_A \int dJ_{1}\, dJ_{2} \, \delta (J_{1} -J_{2}) \, G_\sigma(J_{1})\, G_\sigma(J_{2}) \;
\,e^{-(I[J_1, \psi_1] + I[J_2, \psi_2])}\\
 &=\int \prod_A {\cal D} \psi_A\;  \int dJ\, e^{-\frac{J^2}{2\sigma^2}}\,\, 
\,e^{-(I[J, \psi_1] + I[J, \psi_2])}
 \end{align}
where we have used the rigidity relations \eqref{rigid-G-b-bdys}.
The  final result is the effective path integral for the quantum boundary theory 
\begin{equation}\label{SYKb2-rigidaverage2}
 \langle\,\text{Tr}\, U\left[\kappa_1,  \kappa_2\, ; \,{\cal C}_1\sqcup {\cal C}_2 \right] \rangle =\; \sigma{\sqrt \frac{\pi}{2}}\; \int \prod_A {\cal D} \psi_A\;\, 
\,\;e^{-(I_{eff}[\psi_1] + I_{eff}[\psi_2]) + I_{int}(\psi_1, \psi_2)}
 \end{equation}
where
\be\label{SYKeff}
I_{eff}[\psi] \equiv \,\,\int_{{\cal C}} d\theta\;\;\frac{1}{2}\psi\;  \cdot \; \dot \psi \;\;+ \;\;\frac{\sigma^2}{2}
\int_{{\cal C}} d\theta \int_{{\cal C}} d\theta'\, (\psi(\theta )\cdot\psi(\theta' ))^4 \;, 
\ee
\be\label{SYKeff-int}
I_{int}[\psi_1, \psi_2] \equiv  \;\;\sigma^2 \;
\int_{{\cal C}_1} d\theta_1 \int_{{\cal C}_2} d\theta_2\, (\psi_1 (\theta_1 )\cdot\psi_2 (\theta_2 ))^4
\ee
The case with local insertions $\kappa_1,\kappa_2 \neq 0$ corresponds to add the term $\int_{{\cal C}} d\theta \kappa(\theta) {\cal O}(\theta)\, $ to \eqref{SYK-A} and \eqref{SYKeff}. The action \eqref{SYKeff} stands for an effective SYK theory on each boundary.

Thereby, the main conclusion of this calculation is that two (or $b$) SYK quantum models, with random coupling constants related by a rigidity constraint  \eqref{rigid-G-b-bdys}, behave as two effective SYK models on each boundary, with an effective coupling term between them. Thus, a dominant wormhole solution is able to exist in the dual gravitational theory, and in addition, this effective boundary theory cannot factorize because of the term \eqref{SYKeff-int}.
The (effective) coupling constant is $\sigma^2 \equiv 3! \sigma_0^2 / N^3 > 0$, where the dependence with $N$ turns this coupling negligible in the semi-classical limit.
As shown in ref. \cite{ABS}, in conventional (pure) holography, the presence of a coupling term as \eqref{SYKeff-int} is a necessary condition to have a wormhole dual geometry, while the positiveness of this term is closely related to the possibility of having traversable (or not) dual wormholes \cite{Gao}.

\paragraph{$\bullet$ The fundamental state of SYK$^b$ with holographic JT dual:}

Going deeper into the arguments of Sec \ref{Sec:<SvR-manyb>}, the fundamental state of $b$ independent copies of SYK, namely  SYK$^b$, must be a tensor product of states \eqref{rhogen}
\be
\label{fundamnatalJT} \rho^{\otimes b} = \rho_{\beta_1}(J_1) \otimes \dots \otimes \rho_{\beta_b}(J_b) 
\ee
associated to each disconnected (closed) piece of the boundary. This is because, the holographic dual of the fundamental state shall be a classical solution of \emph{Euclidean} JT gravity\footnote{For further details on this claim, read \cite{VanRaam10} and the discussion related in \cite{Botta2022}.}, but 
\emph{there are no} such Euclidean wormhole solutions connecting two or more boundaries in pure JT gravity. Thus, the correlation functions between different boundaries must vanish, as precisely described by the state \eqref{fundamnatalJT}. In averaged field theories, eq. \eqref{SvR-JT-average-manyb}, the same geometric argument requires that \eqref{GpureJT} is met.

As we will see later, there can be deformations of \emph{pure} JT gravity by including other local fields in the gravity theory, or simply matter fields which can back-react or modify the global structure (topology) of the space-time non trivially. In what follows we will generically 
refer to these theories as \emph{sourced} JT gravity (denoted as sJT), which might admit classical wormholes connecting two (or more) asymptotic boundaries. 
Let us denote the additional bulk local fields by $\chi$, and the back reaction is controlled by the $|| T_{\mu \nu}(\chi) ||$ scale: e.g, if this is much less than $1/G_N$ the back reaction is negligible, and eq. \eqref{GpureJT} approaches the distribution. A non-trivial correction of \eqref{GpureJT} in this scale/parameter, should contain non factorizable terms.

Although \eqref{GpureJT} is a first non-trivial condition on the distribution, there is no general prescription to determine how SYK should be deformed, or its impact on the gravity side.
For example, in Section \ref{Sec:Rigidity-Wh-G2} 
we will consider a particular realization with those features, where the pure JT is deformed with a an imaginary free scalar field, which admits classical wormhole solutions and its dual is suggested to be a particular (rigid) deformation of SYK \cite{Garcia20}.

\subsection{Random (rigid) deformations of SYK and wormholes}
\label{Sec:Rigidity-Wh-G2}

Now consider two disconnected boundaries labeled by $A=1, 2$ and the specific model proposed in \cite{Garcia20} whose phenomenology is suggestively similar to wormhole solutions of certain sJT.
The boundary theory consists of two independent SYK actions \eqref{SYK-A}, defined on each of them. Then it will be deformed by adding the purely imaginary potential
$$ V_A[M] \equiv \,  (-1)^A \; i \kappa \; {\cal O}[M] \,$$ 
 on the respective boundary $A=1,2$. It is similar to the model of Sec \ref{Sec:Orandom1b}: the operator ${\cal O}[M]$ was already defined in \eqref{OpeM} and the source $\kappa$ is taken to be independent on time for simplicity, although in general it can be time dependent.
This situation is slightly different from the framework defined in Sec \ref{Sec:Orandom1b}, since the operator that generates the perturbation is, itself, associated to an independent random coupling $M$.

Let us consider previously the following toy model. The action on each boundary is described by the action 
\be\label{SYK-A-G2}
I[J_A, \Psi_A] \equiv \int_{{\cal C}_A} d\theta\; \left( \frac{1}{2}\psi_A\;  \cdot \;
\dot \psi_A + j_{A; ijkl} \psi_A^i \psi_A^j\psi_A^k\psi_A^l \right)\;
\ee on a closed complex-path ${\cal C}_A$, where $\,j_A\equiv J_A +  (-1)^A i\kappa M_A$.
The lhs of the recipe \eqref{SvR-JT-average-manyb}  in this case reads
\begin{equation}\label{SYKb2-rigidaverage-G2}
\int \prod_A {\cal D} \psi_A \int dJ_{1}\, dJ_{2} \, \, G_\sigma(J_{1})\, G_\sigma(J_{2})  \int dM_{1}\, dM_{2} \, \delta (M_{1} -M_{2}) \, G_\sigma(M_{1})\, G_\sigma(M_{2}) \;
\,e^{-(I[j_1, \psi_1] + I[j_2, \psi_2])}
 \end{equation}
\begin{equation}\nonumber\\ =\int \prod_A {\cal D} \psi_A\;  \int dJ_{1}\, dJ_{2} \, \, G_\sigma(J_{1})\, G_\sigma(J_{2})\int dM\,\,\,  e^{-\frac{M^2}{2 \sigma^2}}\,
\,e^{-(I[J_1 + i\kappa_1  M, \psi_1] + I[J_2 + i\kappa_2  M, \psi_2])}
 \end{equation}
where the rigidity relations \eqref{rigid-G-b-bdys} were assumed only for the (real) coupling constants $M$, and $\kappa_1 =  \kappa_2 \equiv \kappa $ being a \emph{non-vanishing} constant.

Notice that here we did not impose rigidity on the respective SYK coullings $J_1, J_2$ such that they are considered independent; therefore as $\kappa \to 0$, the theory becomes a couple of independent SYK models on each boundary, and then the dual gravity would become strictly JT. However, as argued around eq. \eqref{GpureJT}, the remaining averaging in $K$ is necessary to avoid the factorization paradox, thus in this model, one must demand that $\kappa\neq 0$.

The  final result is the effective path integral for the quantum boundary theory 
\begin{equation}\label{SYKb2-rigidaverage-G22}
 \langle \,\text{Tr}\, U\left[\kappa\, ; \,{\cal C}_1\sqcup {\cal C}_2 \right] \rangle = \sigma{\sqrt \frac{\pi}{2}}\; \int dJ_{1}\, dJ_{2} \, \, G_\sigma(J_{1})\, G_\sigma(J_{2})\int \prod_A {\cal D} \psi_A\;e^{-(I_{}[J_1, \psi_1] + I_{}[J_2, \psi_2]) + I_{int}(\psi_1, \psi_2)}
 \end{equation}
 which consists of two decoupled SYK models, plus an effective potential term 
\be\nn
I_{int}[\psi_1, \psi_2] \equiv  \;\;\frac{(i\kappa)^2\sigma^2}{2}\;\left( \sum_A \int_{{\cal C}_A} d\theta \int_{{\cal C}_A} d\theta'\, (\psi_A(\theta )\cdot\psi_A(\theta' ))^4 -
2\int_{{\cal C}_1} d\theta_1 \int_{{\cal C}_2} d\theta_2\, (\psi_1 (\theta_1 )\cdot\psi_2 (\theta_2 ))^4 \right)
\ee
where the last term is a genuine coupling between the boundaries as expected. It is worth noticing that as $N\to\infty$ this term is negligible, and one recovers two copies of the (exact) SYK/JT correspondence as expected.
Another remarkable fact with this is that now the (effective) coupling constant is $ \kappa^2\; 3! \sigma_0^2 / N^3  \, > 0$ which suggests the traversability of the dual wormhole.
As shown in ref. \cite{ABS}, in conventional (pure) holography, the presence of a coupling term as \eqref{SYKeff-int} is a necessary condition to have a wormhole dual geometry, while the positiveness of this term is closely related to the possibility of having traversable (or not) dual wormholes \cite{Gao}. The model proposed in \cite{Garcia20}, however, consists of imposing the additional rigidity constraint to $J_1, J_2$. This is described by the action 
\be
S[J_A, \Psi_A] \equiv \int_{{\cal C}_A} d\theta\; \left( \frac{1}{2}\psi_A\;  \cdot \;
\dot \psi_A + j_{A ijkl} \psi_A^i \psi_A^j\psi_A^k\psi_A^l \right)\;
\ee on a closed complex-path ${\cal C}_A$, where $\,j_A\equiv J_A +  (-1)^A i\kappa M_A$.
The lhs of the recipe \eqref{SvR-JT-average-manyb} in this case reads
\begin{equation}
\int \prod_A {\cal D} \psi_A \int dJ_{1}\, dJ_{2} \, \delta (J_{1} -J_{2})\, G_\sigma(J_{1})\, G_\sigma(J_{2})  \int dM_{1}\, dM_{2} \, \delta (M_{1} -M_{2}) \, G_\sigma(M_{1})\, G_\sigma(M_{2}) \;
\,e^{-(I[j_1, \psi_1] + I[j_2, \psi_2])}
 \end{equation}
\begin{equation}\nonumber\\ =\int \prod_A {\cal D} \psi_A\;  \int dJ\,\int dM\, e^{-\frac{J^2}{2\sigma^2}}\,\,  e^{-\frac{M^2}{2 \sigma^2}}\,
\,e^{-(I[J + i\kappa_1  M, \psi_1] + I[J + i\kappa_2  M, \psi_2])}
 \end{equation}
where we have used the rigidity relations \eqref{rigid-G-b-bdys} for both real coupling constants $J, M$, and $\kappa_1 =  \kappa_2 \equiv \kappa$ is constant.
The  final result is the effective path integral for the quantum boundary theory 
\begin{equation}
\langle\,\text{Tr}\, U\left[\kappa\, ; \,{\cal C}_1\sqcup {\cal C}_2 \right] \rangle =\; \;  \frac{\pi \sigma^2 }{2}\;\int \prod_A {\cal D} \psi_A\;\, 
\,\;e^{-(I_{eff}[\psi_1] + I_{eff}[\psi_2]) + I_{int}(\psi_1, \psi_2)}
 \end{equation}
where
\be
I_{eff}[\psi]\equiv \,\,\int_{{\cal C}} d\theta\;\;\frac{1}{2}\psi\;  \cdot \; \dot \psi \;\;+ \;\;\frac{( 1 + (i\kappa)^2)\sigma^2}{2}
\int_{{\cal C}} d\theta \int_{{\cal C}} d\theta'\, (\psi(\theta )\cdot\psi(\theta' ))^4 \;, 
\ee
\be
I_{int}[\psi_1, \psi_2] \equiv  \;\;(1 - (i\kappa)^2)\sigma^2 \;
\int_{{\cal C}_1} d\theta_1 \int_{{\cal C}_2} d\theta_2\, (\psi_1 (\theta_1 )\cdot\psi_2 (\theta_2 ))^4
\ee
Now the (effective) coupling constant is $ 3! \sigma_0^2 / N^3 (1+\kappa^2) >0$ \cite{GarciaGarcia2022}.
Notice that as $\kappa \to 0$, one recovers the model \eqref{SYKb2-rigidaverage2}. The dependence of this coupling with $N$ is crucial to get consistency with the requirement \eqref{GpureJT}, since this coupling vanishes in the large $N$ limit, however regarding the duality SYK/JT, the leading boundary theory effectively behaves as a system of free fermions.
In this sense, an interesting modification similar to the previous toy model might be considered by simply relaxing the rigidity in $J$s (eq. \eqref{SYKb2-rigidaverage-G22}). 

Another interesting possibility to be studied is to re-scale the relative coupling, e.g. $\kappa \to 1/N$,
in order to turn the deformation manifestly sub-leading with respect to the coupled SYK model, eq. \eqref{SYKb2-rigidaverage2}.

\subsection{Wormhole correlators}
\label{ConnCorr}

The models discussed so far are suitable deformations of SYK consistent with the averaged holographic formulas (eq. \eqref{SvR-JT-average-manyb}), the large $N$ limit, and are able to describe dual gravity models with wormholes as dominant saddles. In real time, it is particularly interesting to get traversable wormholes. Thereby, the next step is to analyze the existence and properties of these solutions in 2d gravity; namely, sourced JT models that we have defined above.

For multiple boundaries we consider two disconnected SK paths in complex time and assuming the presence of bulk interactions or extra bulk fields that support the wormhole we explore real time gravitational geometries. We take these manifolds as background and compute real time correlation functions for probe scalar fields.
The problem we are set to solve is the two boundary version of \eqref{SvR-JT-average-manyb},
\begin{equation}\label{2boundary-SvR}
    Z_{SYK}\equiv
    \int_{{\cal C}_1;{\cal C}_2}  e^{-I_g-I_{KG}}
\end{equation}
where $\int_{{\cal C}_1;{\cal C}_2}$ represents an integration over all fields with boundary conditions fixed on the asymptotic boundaries defined by the curves ${\cal C}_1,{\cal C}_2$ to be specified below. $I_g$ represents a particular gravitational theory that supports the wormhole as a saddle point. The precise way in which the wormholes are made stable is not of particular importance. For example, one can consider any of the set-ups described in App. \ref{JT-Worm}, for which our geometries would be saddle point solutions. The $I_{KG}$ is taken to be an action for a massive probe scalar $\chi$ 
of which we compute boundary correlators in real time and expectation values on holographic excited states. 

An important comment is due regarding our examples and Hamiltonian choices $H^\pm\equiv H_1 \pm H_2$ in the boundary theories. Recalling our discussion below \eqref{metric}, we have that the prescribed time evolution for a TFD system is generated by $H^- \equiv H_1 - H_2$ \cite{Umezawa} over the SK path in Fig. \ref{Fig:TFD}. The corresponding bulk evolution in this scenario for a single SK boundary involved only the exterior Rindler patches, covered by the time-like Killing vector that does not penetrate into the horizon.
As mentioned in footnote \ref{footHplus}, the complementary $H^+$ scenario for the same SK path was beyond the scope of our work. Essentially it lacks a Global time Killing vector which would make the gravitational dual a complicated complex signature manifold. As a consequence, this scenario was disregarded in Sec. \ref{CorrSec3}.

However, we show that upon considering two SK paths as boundary conditions, a real time saddle associated with the $H^+$ boundary Hamiltonian choice opens up. The $H^+$ scenario will involve an AdS$_2$ version of the real time Thermal AdS scenario \cite{SvRL,us3} or Thermal Wormhole and, as such, segments of Lorentzian Global AdS$_2$ geometry. This time, the $I_g$ wormhole saddle do provide a constant Dilaton profile and the Global time Killing vector is recovered. 
Conversely, the $H^-$ scenario is no longer able to provide a consistent time independent Dilaton profile upon inclusion of the sources and real time segments, thus reversing the situation that we had for the single SK path scenario.

\subsubsection{Thermal wormhole}\label{Thermal}

\begin{figure}[t]\centering
\includegraphics[width=.8\linewidth] {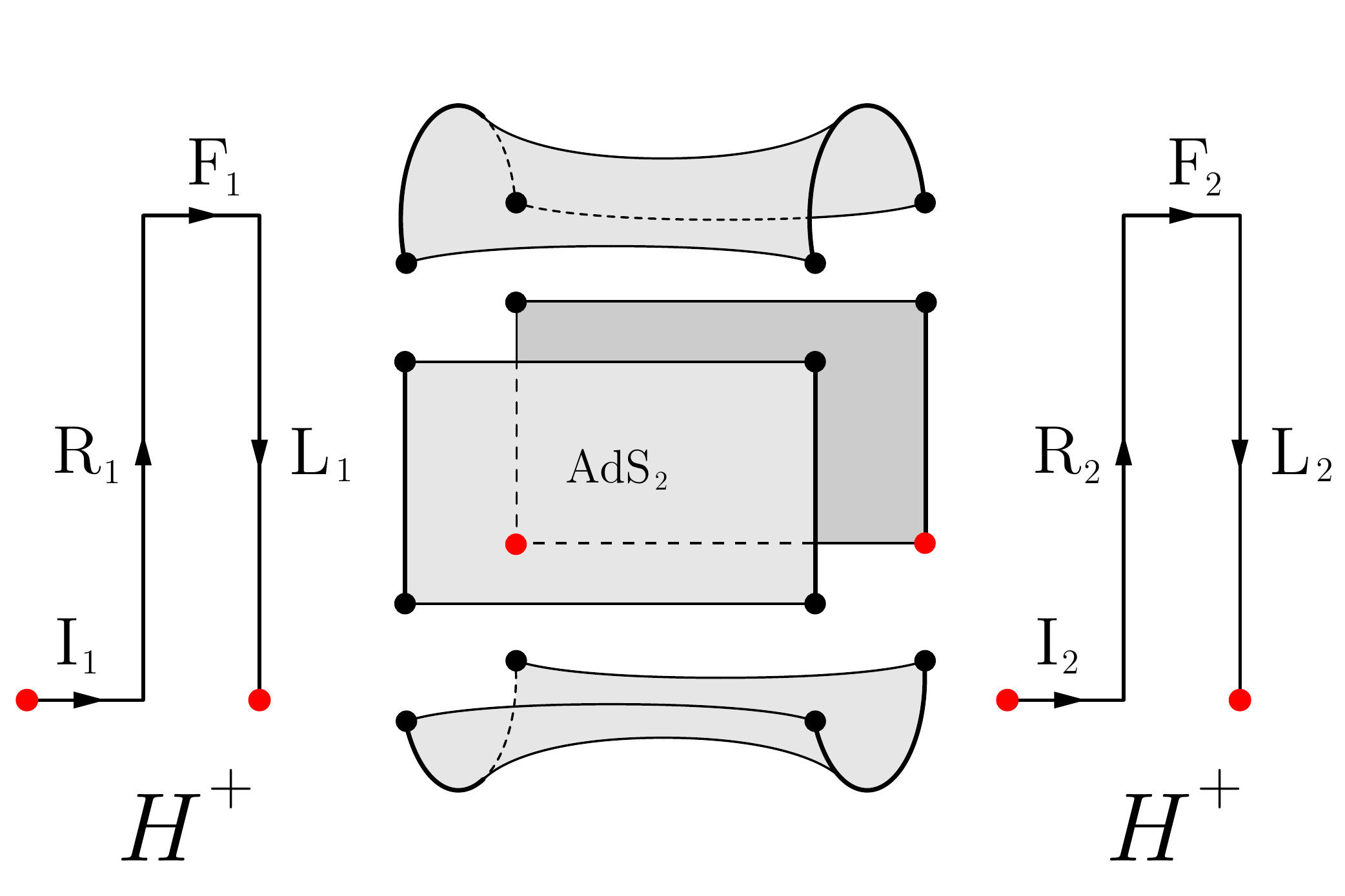}
\caption{On the sides of the figure we show the two SK paths dubbed ${\cal C}_1$ and ${\cal C}_2$. These paths have an effective interaction such that their dynamics are coupled. In the middle of the figure the geometry dual to two SK interacting SK paths is shown. It contains segments of pure Euclidean and Global Lorentzian AdS$_2$.}
\label{Fig:Canaleta}
\end{figure}

We now present the example of a wormhole geometry dual to a couple of Thermal SK path as in Fig. \ref{Fig:Canaleta} with $H^+$ as the boundary Hamiltonian. The resulting bulk saddle is also shown in Fig. \ref{Fig:Canaleta}. The geometry is composed of two Lorentzian and two Euclidean sections of AdS$_2$ of length $\Delta T$ and $- i \beta/2$ respectively. It can be understood as a dimensional reduction of a real time Thermal AdS solution \cite{SvRL,us3}.

Thermal AdS is usually interpreted in higher dimensions as a state of thermal equilibrium between the spacetime and matter, in which the matter is not hot enough to collapse to a Black Hole \cite{HP}. In our scenario, an analogous phase transition occurs at high temperatures to a disconnected geometry \cite{Qi,Garcia20}. Furthermore, there is no exact zero temperature Global AdS$_2$ scenario relevant for holography \cite{Almheiri14,Malda16}.

In any case, as in the higher dimensional thermal AdS scenario, the physical N modes of the geometry retain the pure AdS normal frequencies and its dependence in the temperature appear as a rescaling in the mode normalization. Mathematically, the equations of motion for the probe scalars are identical to the ones in Global AdS$_2$ so the propagating modes must be the same and the topological periodicity can be forced via images method.
We refer to App. \ref{Appendix} for a study of the correlators and N modes of a massive scalar $\chi$ over pure AdS$_2$.

The metrics that cover the geometry are 
\begin{equation}\label{Metric-Canaleta1}
    ds^2=-(r^2+1)dt^2+\frac{dr^2}{r^2+1}\;,\quad ds^2=(r^2+1)d\tau^2+\frac{dr^2}{r^2+1} \;,\quad r,t\in(-\infty,\infty)\quad \tau\in(-\beta/2,\beta/2]
\end{equation}
and the Dilaton profile is
\begin{equation}
    \Phi= \frac{2\phi_b}{\pi} (1+r \arctan(r))
\end{equation}
where $\phi_b$'s exact expression depends on the model that make the wormhole stable, see e.g. \eqref{WHDilatonMQ} and \eqref{WHDilatonGarcia}. We denote the real time boundaries in front of Fig. \ref{Fig:Canaleta} $R_{1},R_{2}$ and the ones in the back as $L_{1},L_{2}$.

We now present scalar real time correlators in the geometry shown in Fig. \ref{Fig:Canaleta} for a Klein Gordon scalar field, see \eqref{Global-EDM} over metric \eqref{Metric-Canaleta1}.
As mandated by the SvR prescription, correlators are computed by: putting boundary sources, solving the equations of motion that meet the gluing conditions, computing the on-shell action and deriving wrt to the external sources to find the correlators using \eqref{2boundary-SvR}. 

The math involved in the computations for this section are highly derivative of those in App. \ref{Appendix} and as such we mostly present results.
To put an example, for a single source $\chi_{R_1}(t)$ source on $R_1$ after solving the equations of motion and meeting the gluing conditions with all other regions one gets, cf. with \eqref{R-NN},
\begin{align}
\chi_{R}(t,r)&=\sum_{n=-\infty}^{\infty} 
\int dt'\int_{{\cal{F}}}\frac{d\omega}{2\pi}e^{-i\omega(t-t'+i n \beta)}q_{1;\omega}(r)\chi_{R_1}(t')\\
&=\frac{ 2^{-\Delta } \Gamma (\Delta)}{\sqrt{\pi } \Gamma \left(\Delta
   -\frac{1}{2}\right)}\sum_{n=-\infty}^{\infty} \int dt' \frac{\chi_{R_1}(t') }{\left(\sqrt{r^2+1}\cos((t-t')(1-i\epsilon)+in \beta)-r\right)^{\Delta}}\;.  
\end{align}
From the many possible correlators available to compute, we are mainly interested in the ones that cross the real time wormhole. To fix notation, we state that the same side $R_1,R_1$ correlator is
\begin{equation}
\langle {\cal O}_{R_1}(t){\cal O}_{R_1}(t')\rangle =\frac{ 2^{-\Delta } \Gamma (\Delta +1)}{\sqrt{\pi } \Gamma \left(\Delta
   -\frac{1}{2}\right)} \sum_{n=-\infty}^{\infty} \frac{1 }{\left(\cos((t-t')(1-i\epsilon)+i n \beta)-1\right)^{\Delta}}\;,    \label{CanCorrRR}
\end{equation}
The correlators that cross from $R_1,R_2$ is
\begin{equation}
\langle {\cal O}_{R_1}(t){\cal O}_{R_2}(t')\rangle =\frac{ 2^{-\Delta } \Gamma (\Delta +1)}{\sqrt{\pi } \Gamma \left(\Delta
   -\frac{1}{2}\right)}   \sum_{n=-\infty}^{\infty} \frac{1 }{\left(\cos((t-t')(1-i\epsilon)+i n \beta)+1\right)^{\Delta}}\;.    \label{CanCorrRL}
\end{equation}
whilst the one that crosses both the wormhole and into the second copy $R_1,L_2$,
\begin{equation}
\langle {\cal O}_{R_1}(t){\cal O}_{L_2}(t')\rangle =  \frac{ 2^{-\Delta } \Gamma (\Delta +1)}{\sqrt{\pi } \Gamma \left(\Delta
   -1/2\right)} \sum_{n=-\infty}^{\infty}  \frac{1 }{\left(\cos(t-t'+i (n+\frac 12) \beta)+1\right)^{\Delta}}\;.    \label{CanCorrRLp}
\end{equation}
The gravitational saddle shows that \eqref{CanCorrRL} must still contain lightcone singularities, but our precise SvR real time prescription mandates that the Lorentzian Thermal correlator is Feynman ordered. On the other hand, \eqref{CanCorrRLp} is no longer required to have a regulator.
Notice interestingly that boundary correlators on the same path such as
\begin{equation}
\langle {\cal O}_{R_1}(t){\cal O}_{L_1}(t')\rangle = \frac{ 2^{-\Delta } \Gamma (\Delta +1)}{\sqrt{\pi } \Gamma \left(\Delta
   -\frac{1}{2}\right)} \sum_{n=-\infty}^{\infty}  \frac{1 }{\left(\cos(t-t'+i (n+\frac 12) \beta)-1\right)^{\Delta}}\;.    \label{CanCorrRRp}
\end{equation}
have also lost its lightcone divergences due to the interaction between paths. One should confront correlators \eqref{CanCorrRR} and \eqref{CanCorrRRp} with the standard matrix correlator for isolated SK paths, see \cite{HerzogSon,us3}. 

As for the expectations values of boundary operators in holographic excited states, one gets that the solution on $R$ for sources in the Euclidean past and future respectively are
\begin{align}
    \chi_{I}( t,r)&=\sum_{m=-\infty}^\infty \sum_{n=0}^\infty e^{-i(\Delta+ n) (t+i m \beta)} \left(\bar\chi_{I_{1}}(\Delta + n ) + \bar\chi_{I_{2}}(\Delta + n)\right) \mathop{\mathrm{Res}}_{\omega=\Delta+n}f_L(\omega,r)
\end{align}
\begin{align}
\chi_{F}( t,r)&=\sum_{m=-\infty}^\infty\sum_{n=0}^\infty e^{+i(\Delta+ n)  (t+i m \beta)} \left(\bar\chi_{F_{1}}(\Delta + n ) + \bar\chi_{F_{2}}(\Delta + n)\right) \mathop{\mathrm{Res}}_{\omega=\Delta+n}f_L(\omega,r) 
\end{align}
where we have denoted,
\begin{equation}
    \bar\chi_{I_{1/2}}(\omega)=\int_{-\frac \beta2}^{0} d\tau e^{\omega \tau}\phi_{I_{1/2}}(\tau) \qquad\qquad \bar\chi_{F_{1/2}}(\omega)=\int^{\frac \beta2}_{0} d\tau e^{-\omega \tau}\phi_{F_{1/2}}(\tau)
\end{equation}
and $\phi_{I_{1/2}}(\tau)$ and $\phi_{F_{1/2}}(\tau)$ are the Euclidean sources on the I and F regions, defined to be non-trival only on $\tau\in[-\beta/2,0]$ and $\tau\in[0,\beta/2]$ respectively.
The expectation value of boundary operators can now be obtained via the BDHM dictionary, i.e.
\begin{align}
    \langle \chi_F | {\cal O}_R(t)| \chi_I\rangle 
    &\equiv (2\Delta-1)\lim_{r\to+\infty} r^{\Delta} \left(\chi_I( t,r)+\chi_F( t,r)\right) \\\nn
    &=\sum_{m=-\infty}^\infty\sum_{n=0}^\infty \left[
    e^{-i(\Delta+ n) t} \left(\bar\chi_{I_{1}}(\Delta + n ) +\bar\chi_{I_{2}}(\Delta + n ) \right)\right.\\
    &\qquad\qquad\qquad\left.
    +e^{+i(\Delta+ n) t} \left(\bar\chi_{F_{1}}(\Delta + n ) +\bar\chi_{F_{2}}(\Delta + n ) \right) \right] F^\beta_{\Delta,n}\label{CanEVR}\\
    &=\sum_{m=-\infty}^\infty  \left( \int_{-\frac \beta2}^{\frac \beta2} d\tau \frac{\chi_{I_{1}}(\tau)+\chi_{F_{1}}(\tau)}{\left[ \cos(t+i\tau-i m \beta)-1 \right]^{2\Delta}} + \int_{-\frac \beta2}^{\frac \beta2} d\tau \frac{\chi_{I_{2}}(\tau)+\chi_{F_{2}}(\tau)}{\left[ \cos(t+i\tau-i m \beta)+1 \right]^{2\Delta}}\right) \label{CanEVRresum}
    \end{align}
where we have written the ressumed expression in \eqref{CanEVRresum} to make explicit that the sum of modes in $n$ makes the expression convergent. Our main result is \eqref{CanEVR}, where the propagating modes of the solution are made explicit and the inherited mode normalization from holography in this finite temperature scenario is,
\begin{align}\label{CanEVModeNorm}
   F^\beta_{\Delta,n}&= 
   \frac{ \Gamma \left(\frac{3}{2}-\Delta \right)
   \Gamma (n+2 \Delta )}{4^{\Delta-1 }\pi  \Gamma \left(\Delta
   -\frac{1}{2}\right) \Gamma (n+1)}e^{-(\Delta+n) m \beta}\;.
\end{align}
An analogous result for $ \langle \chi_F | {\cal O}_{L_1}(t)| \chi_I\rangle $ can be found upon exchanging 1$\leftrightarrow$2 in \eqref{CanEVR}. We must emphasize that this set of correlators, expectation values and their attached geometry Fig. \ref{Fig:Canaleta} dual to a couple of Schwinger Keldysh paths, alongside their precise time-orderings can be taken as a nice corollary of our work.

\section{Conclusions}
\label{sec:conclusiones}

In this paper we have extended the Skenderis and van Rees 
 and the BDHM prescription for real time holography in the context of the JT/SYK correspondence, and explored its consequences on both the factorization problem and the role of the ensemble averaging in SYK. 

In doing so, we were able to show a number of novel properties of real time holography in the JT/SYK correspondence. For example, topological arguments show that for the holographic gravitational problem to be consistent in 1+1 dimensions, the SK path defining the QM theory in the boundary must be closed. 
Put in other words, the dynamics of holography for a single path in JT/SYK is always exploring physics of a thermal initial state or perturbations of such a state. 
In particular, we should highlight eq. \eqref{SvR-JT-average-manyb} as an improved $\langle$SvR$\rangle$ prescription which corresponds to a unified equation capturing the holographic real time prescriptions and average ensemble for an arbitrary number of boundaries. By construction, the prescription in eq. \eqref{SvR-JT-average-manyb} is free of the factorization problem and the role of ensemble averaging is explored.
Among other things, we also revisited well-established mechanisms to construct traversable wormholes \cite{ABS,Gao,Qi} in this context and updated the criteria defined in \cite{ABS}. 

To illustrate our construction, we presented a couple of relevant examples both for single and multiple boundary SK paths. The first example provides a real time saddle point solution dual to a single SK path. The geometry and its phenomenology can be understood as a dimensional reduction of the solution found in \cite{us3,us4} and shows that there are real time scenarios that hold from traditional to JT/SYK holography.
The second example presented is a novel real time holographic dual to a couple of either entangled or interacting theories defined over two SK paths. The complete manifold is seen to contain segments of real time Global AdS$_2$ (it can be seen as a dimensional reduction of Thermal AdS$_2$) and is checked to be relevant in the context of (modified) JT/SYK holography. Thus, the canonically quantized modes travelling the system are the same as in Global AdS$_2$, adequately rescaled by the temperature. Explicit results for correlators and expectation values of boundary operators over holographic excited states are also studied and a number of interesting properties are presented. Most notably, it can be seen that the effective coupling between the disconnected boundaries modifies the correlation between real time segments on the same path.
Interestingly, the examples also serve to illustrate an important choice on boundary time $H^\pm=H_R\pm H_L$ which we have not seen emphasized elsewhere in the literature.

As for future directions, it would be interesting to find further novel connected solutions for a $b>1$ set of boundaries and study the correlations that such a geometry implies between the theories. If done systematically, one could in principle try to derive constraints on the $G(J_1,\dots,J_b)$ that define the ensemble average on the QFT side of the duality, see e.g. \eqref{GpureJT}. We leave this possibility for future work.

\section*{Acknowledgments}

We thank J.Russo for relevant feedback on this work.
RA is supported by UNLP and CONICET grant PIBAA (2022 - 2023),  MBC is supported by UNLP and CONICET grants X791, PIP 2017-1109 and PUE Búsqueda de nueva Física. PJM is supported by CONICET, CNEA, and Universidad Nacional de Cuyo, Argentina.

\appendix

\section{Some realizations of wormholes in sourced JT }
\label{JT-Worm}

The models discussed in \ref{Sec:Rigidity-Wh-G2} show that there are suitable deformations of SYK, that because of fundamental requirements of consistency of the holographic formulas (e.g. factorization), are able to describe dual gravity models with wormholes as dominant saddles. In real time $1+1$ gravity, it is particularly interesting to study traversable wormholes. 
This is mostly due to its mathematical simplicity that allows for explicit solutions. We now present two realizations of these wormholes studied in detail in \cite{Qi} and \cite{Garcia20} that provide physical context to the correlator computation that we will be computing in this section below. In particular for \cite{Garcia20} we will be able to provide some physical insight to some expectation values computations that were unclear before.

In the context of our present work and previous criteria on existence of wormhole solutions \cite{ABS}, we can classify these solutions as being dual to:
\textbf{a)} an explicit coupling between operators on disconnected boundaries in pure holography, or \textbf{b)} averaging over random couplings with a rigid constraint.

\subsection{Explicit coupling between operators on disconnected boundaries}

A natural set-up to realize wormhole solutions was the approach studied in \cite{Qi} on which the Gao-Yafferis-Wall mechanism \cite{Gao} in higher dimensions was adapted to a JT scenario. The method consists on explicitly coupling the QM dual theories on each side in a particular way such that an effective negative energy density arises in the bulk interior. This opens up a window of time in which the wormhole between the causally disconnected theories become connected. 
To be concrete, one adds the interaction $g \sum_i \int du\; {\cal O}^i_1(u){\cal O}^i_2(u)$, $g>0$ to \eqref{JT}, where ${\cal O}^i_{1/2}(u)$, $i=1,\dots,N$, are $N$ boundary operators dual to free fields in AdS corresponding to the 1 and 2 boundaries, all with conformal dimension $\Delta$. One then takes a saddle point approximation on the effect of the interaction, i.e.
\begin{equation}
    \langle e^{i g \sum_i^N \int du\; {\cal O}^i_1(u){\cal O}^i_2(u) } \rangle \sim e^{ i g \sum_i \int du\; \langle {\cal O}_1 {\cal O}_2 \rangle } \;,
\end{equation}
with $t_{1/2}(u)$ on their own have Schwarzian actions. Since all operators in the boundary interaction share conformal dimension $\Delta $, the sum over $i$ ends up putting an extra $N$ in front of the interaction, enhancing its effect and making $g N$ the effective coupling constant, which is kept fixed as $N\to\infty$ and $g\to0$. As anticipated, this interacting action leads to equations of motion that allow for a traversable wormholes as a saddle point with Dilaton profile,
\be \label{WHDilatonMQ}
\Phi=
 g N^{2\Delta} \Big(1+ r \arctan(r)  \Big)
\ee
The throat size is linear in the interaction's coupling constant $g>0$. 
In \eqref{WHDilatonMQ} we have disregarded a Casimir energy contribution, which will always be subleading in the parameter regimes we are interested in.
This wormhole is presented directly as a non-local relevant deformation of the quantum mechanics action and as such can be immediately analytically extended to a generic SK closed contour $\cal C$.

\subsection{Averaging over random couplings with a rigid constraint}

A less standard approach within sJT to produce Euclidean wormhole as saddle point solutions to the Euclidean version of \eqref{JT} was presented in \cite{Garcia20}. 
The authors study a system of two non-interacting SYK theories but impose a rigid constraint between the (complex) couplings of both theories, which leads to an effective coupling between them after averaging. They find phases of this SYK system that are reminiscent to wormholes on the gravity side. 

An effective gravitational dual model in JT is proposed by the authors. A marginal ($\Delta=1$) scalar is added to the JT action with pure imaginary boundary sources. We choose $\xi$ rather than $\chi$ for this particular scalar field since it would be taken as part of the gravity theory rather than a probe field over a solution. This coupled gravity equations of motion are solved exactly and the imaginary sources provide the effective negative energy density in the bulk to support the Euclidean wormhole. We will focus on this effective gravity model regardless of the precise SYK dual.

The extra term added to the action in \eqref{JT} is
\begin{equation}\label{action-chi}
    I_\xi = \frac12 \int \!\!\sqrt{g} \;\partial_\mu \xi\partial^\mu \xi,
\end{equation}
with $\xi\to (-1)^A i \kappa$ at the asymptotic boundaries $A=1,2$. 
The boundary conditions are pure imaginary sources $\pm i \kappa $, $\kappa\in\mathbb{R}$ at the asymptotic boundaries given by thermal cycles ${\cal C}_{1,2}$, each of physical length $\beta$, c.f. with Sec. \ref{Sec:Rigidity-Wh-G2}.
The coupled equations of motions derived from these actions can be consistently solved by
\begin{equation}\label{Glob-Metric}
    ds^2=(r^2+1)d\tau^2+\frac{dr^2}{r^2+1},\qquad \tau\in[0,\beta)\qquad r\in(-\infty,\infty)
\end{equation}
where $r\to\mp\infty$ are identified with $1,2$ boundaries respectively. 
The EOMs for $\xi$ are standard KG equations on fixed AdS$_2$, with a unique solution that meets the boundary conditions \eqref{action-chi}
\begin{equation}\label{GEOMs1}
    \square\xi=0 \qquad\Rightarrow\qquad \xi=\frac{2i\kappa}{\pi}\arctan(r)
\end{equation}
The metric equations of motion, now coupled to the massless scalar, 
provide a differential equation for $\Phi$
whose solution can be written as
\be \label{WHDilatonGarcia}
\Phi=\frac{2 \kappa ^2}{\pi ^2}\Big(1+ r \arctan(r)  \Big)\;,
\ee
We now describe some of its properties. The most important role is played by $\kappa\in\mathbb{R}$ providing a net positive contribution for $\Phi$ at $r=0$ opening up the wormhole. We stress that this is an exact solution to the full nonlinearly coupled JT+$\xi$. However, the wormhole geometry as a saddle depends strongly on two conditions, i.e. $\xi$ being massless and the sources being imaginary. As in \eqref{WHDilatonMQ}, in eq. \eqref{WHDilatonGarcia} we have dropped a contribution which is subleading in our regime of interest \cite{Garcia20}. Classical solution \eqref{WHDilatonGarcia} should be identified with the $\chi_0$ classical solution in the context of eq. \eqref{Qcanonical}.

\subsubsection{A comment on imaginary sources}

\begin{figure}[t]\centering
\begin{subfigure}{0.49\textwidth}\centering
\includegraphics[width=.9\linewidth] {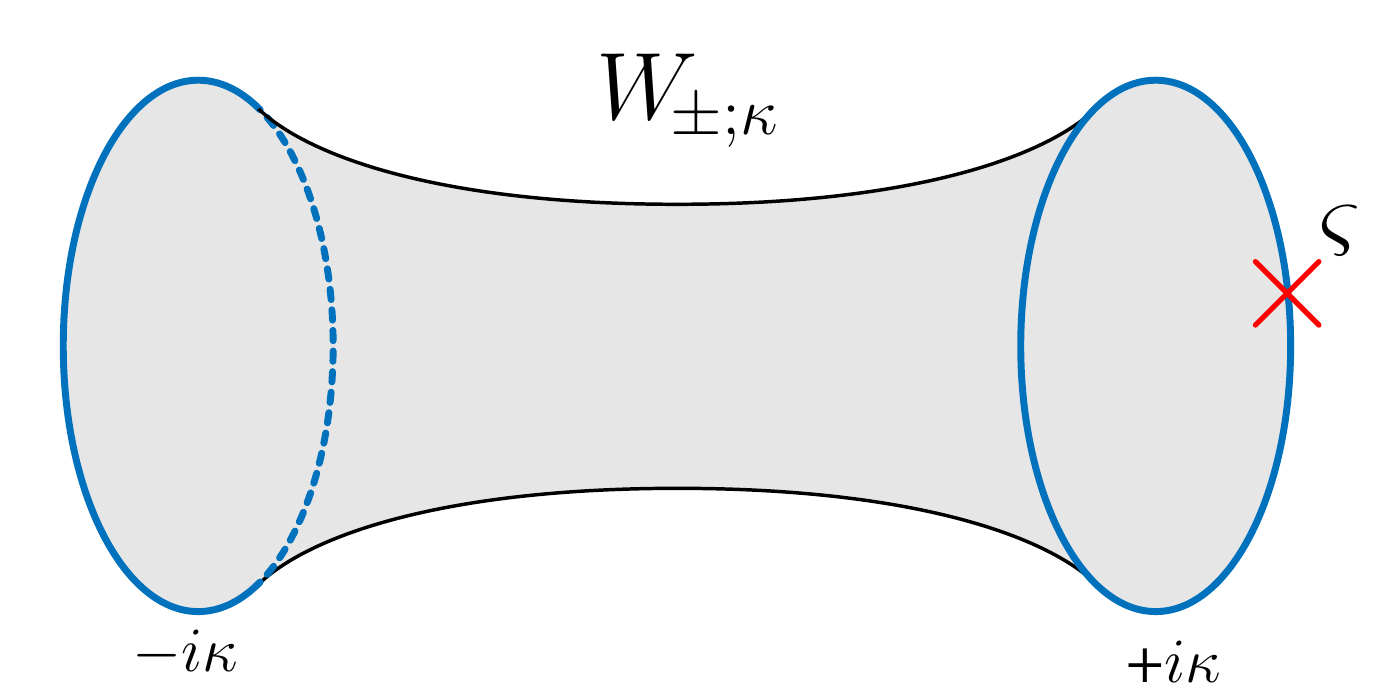}
\caption{}
\end{subfigure}
\begin{subfigure}{0.49\textwidth}\centering
\includegraphics[width=.9\linewidth] {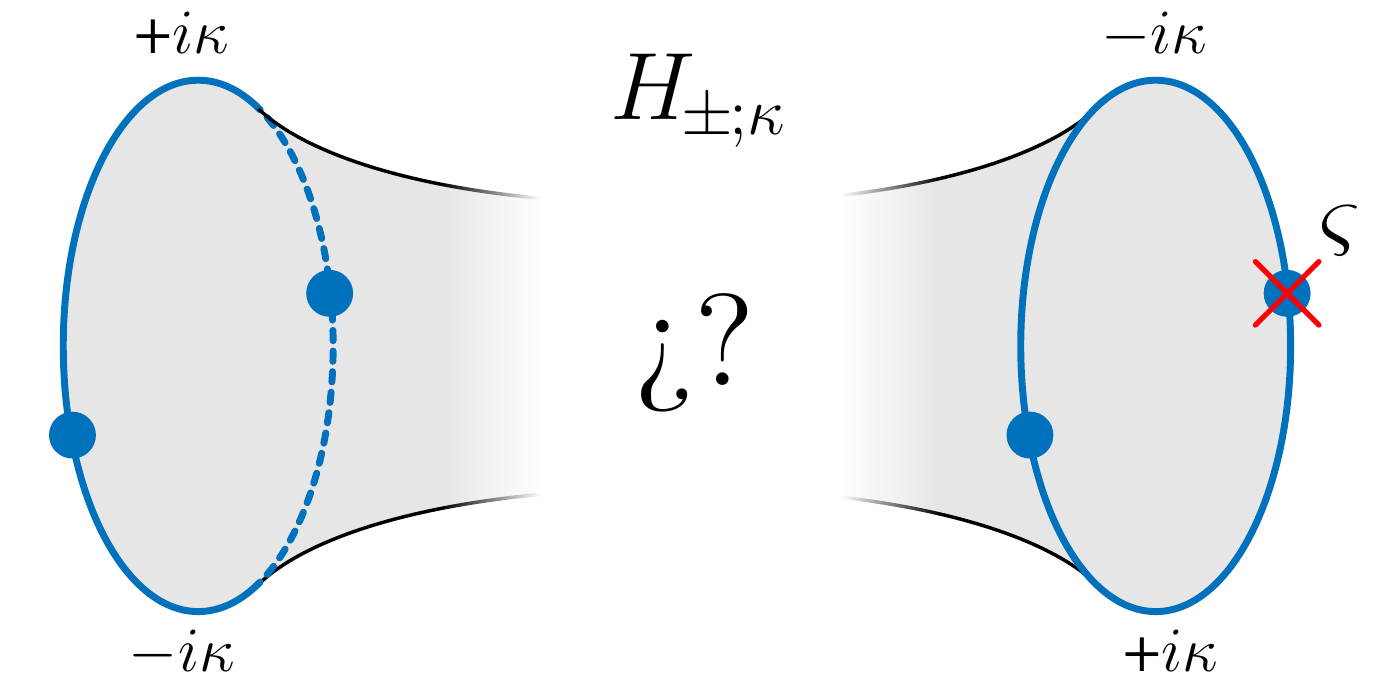}
\caption{}
\end{subfigure}
\caption{(a) A graphical representation of $P_{\pm;\kappa}$ is presented (b) Taking $\xi$ as a Hermitian operator upon quantization, an Euclidean Time Reflection operation leads to a second object $H_{\pm;\kappa}$ to provide the $\xi$ field expectation value. A saddle point for $H_{\pm;\kappa}$ is unknown.}
\label{Fig:GState}
\end{figure}

To conclude this Appendix, we briefly comment on imaginary sources for real scalar fields in the light of the discussion in Sec. \ref{Sec:HES}. 
A property of this wormhole solution due to \eqref{GEOMs1} is that the imaginary sources induce an imaginary result for the path integral
\begin{equation}\label{GVEV}
  W_{\pm;\kappa}= \lim_{r\to\pm\infty}\;\;\int_{\pm i \kappa} {\cal D}\xi \; \xi \; e^{-I_{JT}-I_{\xi}}= \pm i\kappa
\end{equation}
in which integration over the gravitational field is implicit and which in \cite{Garcia20} is interpreted as an imaginary expectation value for the boundary operators. In the language of Sec. \ref{TFDpurification} this can be rewritten as
\be\label{expimaginary}
W_{\pm;\kappa}= \lim_{r\to\pm\infty}\;\;\text{Tr}\{ U_{+i\kappa;-i\kappa} \,\xi\,  U_{+i\kappa;-i\kappa} \}=\lim_{r\to\pm\infty}\;\;\text{Tr}\{ U_{+i\kappa;-i\kappa}\,U_{+i\kappa;-i\kappa} \,\xi \} =\pm i\kappa
\ee
where $U_{+i\kappa;-i\kappa}$ is defined as the Hartle-Hawking gravity wave-functional preparing an initial state with boundary conditions $\pm i\kappa$ on the boundaries, see Fig. \ref{Fig:GState}(a).
Now, considering the pure imaginary result, it cannot be the case that $U_{+i\kappa;-i\kappa}\,U_{+i\kappa;-i\kappa} $ and $\xi$ are both Hermitian operators. We have two choices. Either \textbf{a)} $\xi$ is canonically quantized as an anti-Hermitian operator, see eqs \eqref{Qcanonical} \eqref{BDHM2-ave}). This amounts to rescale $\xi \to i \xi$ off shell and reinterpret this construction as a saddle with real fields but an overall minus sign in the action of the scalar eq. \eqref{action-chi}. From this perspective, the appearance of a wormhole saddle is not entirely surprising as one is coupling exotic negative energy matter fields to gravity, see e.g. \cite{Dimitrov:2021csq}. Option \textbf{b)} is that the field $\xi$ is kept Hermitian, and the $\pm i \kappa$ sources are retained as imaginary sources for a real field. Then, a careful Euclidean conjugation of $U_{+i\kappa;-i\kappa}$ \cite{Jackiw}, implies a reinterpretation of \eqref{GVEV} as a matrix element of $\xi$ rather than an expectation value, which would also explain a complex result in the computation. This would be more in line with the general $d>1$ discussion in Sec. \ref{Sec:HES}.
In this language, the expectation value of the operator requires to reflect the sources on the same boundary, and a Hermitian state is defined as, see Fig. \ref{Fig:GState}(b),
\be
H_{\pm;\kappa}\equiv
\lim_{r\to\pm\infty}\;\;\text{Tr}\{U_{+i\kappa;-i\kappa}\,U_{-i\kappa;+i\kappa}  \,\xi \}
\qquad\qquad (U_{+i\kappa;-i\kappa}\,U_{-i\kappa;+i\kappa})^\dagger=(U_{+i\kappa;-i\kappa}\,U_{-i\kappa;+i\kappa})
\ee
Notice that this is a different computation than $P_{\pm;\kappa}$. Based on our previous developments reviewed in Sec. \ref{Sec:HES} one would expect that
\begin{equation}\label{GTVEVs}
   \lim_{r\to\pm\infty}\;\;\text{Tr}\{U_{+i\kappa;-i\kappa}\,U_{-i\kappa;+i\kappa}  \,\xi \}=0 \qquad\qquad \lim_{r\to\pm\infty}\;\;\text{Tr}\{U_{+i\kappa;-i\kappa}\,U_{-i\kappa;+i\kappa}  \,\Pi_\xi \} \propto \pm \kappa
\end{equation}
resembling \eqref{bcEOM}, where $\Pi_\xi$ is taken to be the conjugated momentum to the bulk field $\xi$. Notice, however, that we are not claiming that a wormhole saddle (or any other geometric saddle) exists for the boundary conditions in \eqref{GTVEVs}. This is emphasized in Fig. \ref{Fig:GState}(b). Furthermore, the abrupt sign flip $+i\kappa\to-i\kappa$ on each boundary should require high frequency modes for $\xi$, and thus a novel (and time dependent) solution for the non-linearly coupled JT+$\xi$ equations of motion, which is beyond the scope of this work. We stress that \eqref{GTVEVs} must be taken as a conjecture based on higher dimensional results and not as an explicit computation. We also point out that this analysis was entirely made from a gravitational JT+$\xi$ point of view. We leave a study of the consequences of this conjecture from the QFT side for future work.

We conclude by a making a final comment on this construction. It is interesting to note that $U_{+i\kappa;-i\kappa}$ interpreted as an holographic excited state with imaginary sources would be the first case in which the coupled gravity + matter equations of motion were exactly solved and the $\kappa\neq0$ state is able to provide a different topology than the $\kappa=0$ case. This would be the first non-perturbative study in this regard. It would be interesting to consider higher dimensional analogues of this mechanism.

\section{Scalar field in pure AdS$_2$}
\label{Appendix}

In this App. present the relevant AdS$_2$ coordinates for the main body of the text and some relevant classical solutions for a massive scalar field $\chi$ over pure Global AdS$_2$. We review why this set-up on its own is not adequate for holography in 1+1 dimensions, but it is still useful for our purposes on the light of Sec. \eqref{Thermal}.  
 
\begin{figure}[t]\centering
\includegraphics[width=.9\linewidth] {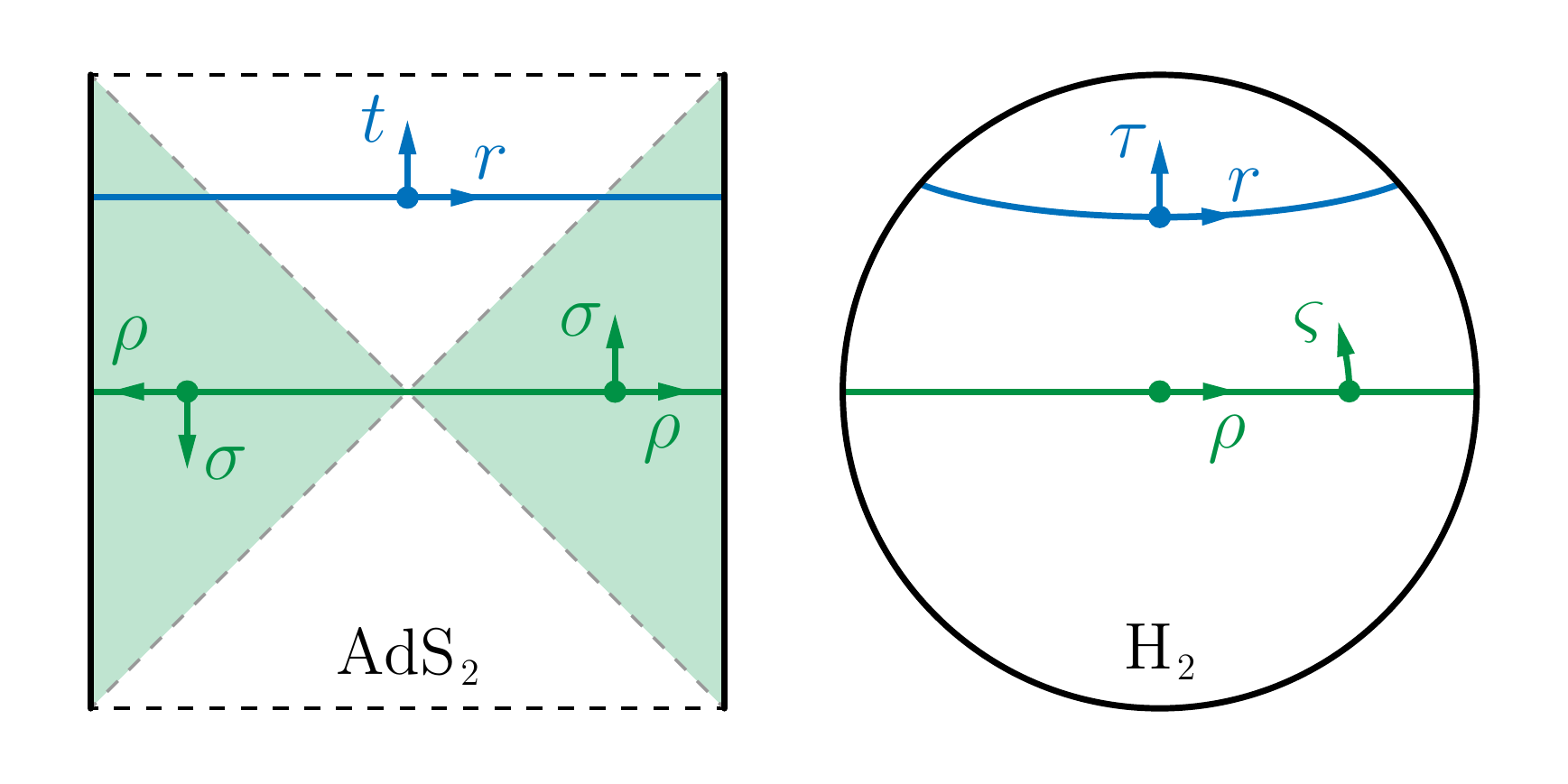}
\caption{The relevant foliations used in this work are shown both for Lorentzian AdS$_2$ and Euclidean AdS$_2$, i.e. H$_2$. The coordinates are defined in \eqref{Fol}.}
\label{Fig:Fol}
\end{figure}

We define AdS$_2$ as a hypersurface defined over $1+2$ flat spacetime
\begin{equation*}
    -U^2-V^2+X^2=-1\qquad ds^2=-dU^2-dV^2+dX^2\;,
\end{equation*}
and define the two foliations,
\begin{equation}\label{Fol}
    U=\sqrt{r^2+1}\cos(t)=\rho\qquad V=\sqrt{r^2+1}\sin(t)=\sqrt{\rho^2-1}\sinh(\sigma)\qquad X=r=\sqrt{\rho^2-1}\cosh(\nu)
\end{equation}
We will also use the Euclidean analytic extensions of these metrics $t\to-i\tau$ and $\sigma\to-i\varsigma$.
We show how these coordinates cover the AdS$_2$ and Euclidean AdS (H$_2$) manifolds in Fig. \ref{Fig:Fol}.

For this Appendix we need the Global Lorentzian AdS$_2$ foliation in terms of $\{r,t\}$ in \eqref{Fol},
\begin{equation}\label{App:Lmetric}
    ds^2=-(r^2+1)dt^2+\frac{dr^2}{r^2+1},\qquad r,t\in(-\infty,\infty)
\end{equation}

A solution of the JT EOMs, see eq. \eqref{JT}, requires also a  Dilaton $\Phi$ profile on top of the metric \eqref{App:Lmetric}.
Profiting from examples that reduce to JT upon dimensional reduction, it is customary to associate the magnitude of the Dilaton $\Phi$ as the size of the wormhole at any given Cauchy slice, see e.g. \cite{Malda16}. Thus, Global AdS$_2$ (an eternally traversable wormhole) should also support a non-trivial time independent Dilaton solution to be considered relevant for the JT/SYK correspondence.
It can be shown that there exists no such Dilaton solution
on pure Global AdS$_2$ \cite{Qi}.

However, we found in Sec. \ref{Thermal} that the saddle point geometry dual to two disconnected SK paths yields a manifold saddle that can be understood as a Thermal AdS$_2$ solution, over which we aim to compute probe scalar correlators. We thus find useful to study a massive scalar KG field in Global AdS$_2$, to introduce the relevant notation and present useful analysis. We will denote for future convenience the two AdS$_{1+1}$ boundaries as $R_1$ and $R_2$ and also retain standard holography notation such as $\langle{\cal{O}}_{R_1}(t){\cal{O}}_{R_1}(t')\rangle$ and $\langle{\cal{O}}_{R_1}(t){\cal{O}}_{R_2}(t')\rangle$ for AdS boundary correlators even if there is no known realizations of a QM ${\cal O}_{R_{1/2}}(t)$ operators dual to only pure AdS$_2$ for they will become physical in the Thermal AdS$_2$ scenario.

\subsection{Real-time correlation functions}

We take a probe scalar, $\chi(t,r)$ in the Lorentzian metric \eqref{App:Lmetric} with action
\begin{equation}\label{Global-EDM}
    I_{KG} = -\int \sqrt{g}(\partial_\mu \chi \partial^\mu \chi+m^2 \chi^2)\:,\qquad\qquad \left(\square-m^2\right)\chi=0\,,
\end{equation}
One can expand the solution in a Fourier basis $\chi(t,r)=e^{i \omega t}q(r)$ to obtain the radial equation,
\begin{equation}
  \left((r^2+1)q'(r)\right)'=\left(\Delta(\Delta-1)-\frac{\omega^2}{r^2+1}\right)q(r)
\end{equation}
with $m^2=\Delta(\Delta-1)$. A solution to this equation is
\be 
q_{1;\omega}(r)=C_1 P_{\Delta-1}^\omega(i r)+C_2 Q_{\Delta-1}^\omega(i r)\,,\qquad C_2=\frac{2 C_1}{ i \pi} \,,\qquad
C_1=\frac{i \sqrt{\pi }  e^{-\frac{1}{2} i \pi 
   (\Delta +\omega )} \Gamma (\Delta -\omega )}{2^{\Delta }\Gamma
   \left(\Delta -\frac{1}{2}\right)}\,,
\ee
where $P_{a}^{b}(x)$ and $Q_{a}^{b}(x)$ are the Legendre functions of the first and second type respectively. The coefficients are chosen so that the solution diverges at $R_1$, $r\to+\infty$ and is regular at $R_1$, $r\to-\infty$, i.e.
\begin{align}
    q_{1;\omega}(r)&\sim 1 \times r^{\Delta-1}+{\cal O}(r^{-\Delta})
   \,,\qquad\qquad r\rightarrow\infty,\\
   q_{1;\omega}(r)&\sim {\cal O}(r^{-\Delta})
    \,,\qquad\qquad\qquad\qquad r\rightarrow-\infty.
   \label{leftexp}
\end{align}
As defined, $q_{1;\omega}$ has poles in the complex $\omega$ plane at $\omega=\pm (\Delta+n)$, where $n\in \mathbb{N}$. With $q_{1;\omega}(r)$ at hand, one can define a solution with boundary conditions $\chi(t,r)\sim r^{\Delta-1} \chi_{R_1}(t)$ at $r\to+\infty$ as
\be 
\chi_R(t,r)=\int dt'\int_{{\cal{F}}}\frac{d\omega}{2\pi}e^{-i\omega(t-t')}q_{1;\omega}(r)\chi_{R_1}(t')
=\frac{ 2^{-\Delta } \Gamma (\Delta)}{\sqrt{\pi } \Gamma \left(\Delta
   -\frac{1}{2}\right)} \frac{1 }{\left(\sqrt{r^2+1}\cos((t-t')(1-i\epsilon))-r\right)^{\Delta}}\;,\label{R-NN}
\ee
where the sub-index ${\cal{F}}$ means that we have chosen a Feynman path for the $\omega$ integral. An analogous solution with sources at $r\to-\infty$ can be defined as $\chi_R(t,r) \to \chi_R(t,-r)$.

By looking at the speciffic frequencies $\omega=\pm (\Delta+n)$ one finds that there also exist an infinite set of solutions that decay at both $r\to\pm\infty$, i.e. the AdS$_2$ N modes. These can be added to any real time solution without altering the asymptotic boundary conditions. The most general set of these modes can be written,
\be \label{App:NModes}
q_N(t,r)=\sum_{n=0}^{\infty}\left(C_n e^{-i(\Delta+t)n} \mathop{\mathrm{Res}}_{\omega=\Delta+n}q_{1,\omega}(r)+D_n e^{i(\Delta+t)n} \mathop{\mathrm{Res}}_{\omega=\Delta+n}q_{1,\omega}^*(r)\right).
\ee
We will find these modes of use for computing expectation values on excited states below.

The Feynman correlator for the theory on the $R_1$ boundary can be obtained by considering the on-shell action in \eqref{Global-EDM} over a solution consisting of \eqref{R-NN} and with all N modes in \eqref{App:NModes} turned off. This yields
\begin{equation}\label{App:S0RR}
I^0_{11}\equiv 
-\frac{i}{2}\int dt\,\,dt'\chi_{R_1}(t)K_{11}(t,t')\chi_{R_1}(t')
\end{equation}
where the kernel can be seen to be the time ordered correlator $\langle{\cal{O}}_{R_1}(t){\cal{O}}_{R_1}(t')\rangle$, i.e.
\begin{align}
    \langle{\cal{O}}_{R_1}(t){\cal{O}}_{R_1}(t')\rangle&=-i K_{11}(t,t') \nn\\ &=-i r^\Delta\int_{{\cal F}}
 \frac{d\omega}{2\pi} e^{-i\omega(t-t')} \left( r\partial_r q_{1,\omega(r)}\right)_{r\rightarrow\infty} \nn \\
    &=-i\frac \Delta2 \int_{{\cal{F}}} d\omega\; e^{-i\omega(t-t')}
   \times \\
   &\qquad\qquad\qquad\times \frac{ \Gamma
   \left(\frac{1}{2}-\Delta \right)
   \left(\csc (\pi  \omega ) \csc (\pi 
   (\Delta +\omega ))-e^{i \pi  \omega }
   (\cot (\pi  \omega )+i) \csc (\pi  (\Delta
   -\omega ))\right)}{4^{ \Delta }e^{i \pi 
   (\Delta +\omega )}\Gamma \left(\Delta
   -\frac{1}{2}\right) \Gamma (1-\Delta
   -\omega) \Gamma (1-\Delta +\omega)} \nn \\
   &= \frac{ 2^{-\Delta } \Gamma (\Delta +1)}{\sqrt{\pi } \Gamma \left(\Delta
   -\frac{1}{2}\right)} \frac{1 }{\left(\cos((t-t')(1-i\epsilon))-1\right)^{\Delta}}\;.\label{AdS2G-RR}
\end{align} 
In the last line we noted that the integral closes differently for $(t-t')\gtrless0$, capturing the residues at $\omega=\pm(\Delta+n)$ respectively. A similar analysis to the one around eq. \eqref{onchell} shows that the integral expression is equivalent the Feynman order correlator, as written in the last line.

To give an example, we can explicitly carry the integral for $(t-t')<0$ capturing the $\omega=+(\Delta+n)$, 
\begin{align}
   \langle{\cal{O}}_{R_1}(t){\cal{O}}_{R_1}(t')\rangle
   &=\frac{  \Delta   (\cot (\pi  \Delta )+i)
   \Gamma \left(\frac{1}{2}-\Delta \right) }{4^{\Delta }\Gamma \left(\Delta
   -\frac{1}{2}\right) }\sum_{n=0}^{\infty} \frac{(-1)^n e^{-i (n
   t+\Delta  (t+\pi ))} }{\Gamma (n+1) \Gamma (1-n-2 \Delta)} \\
   &=\frac{ 2^{-\Delta } \Gamma (\Delta +1)}{\sqrt{\pi } \Gamma \left(\Delta
   -\frac{1}{2}\right)} \frac{1 }{\left(\cos(t-t')-1\right)^{\Delta}}\;.
\end{align} 
The correlators $\langle{\cal{O}}_{R_2}(t){\cal{O}}_{R_2}(t')\rangle$ on $r\to-\infty$ can be seen to be also time ordered and identical to $\langle{\cal{O}}_{R_1}(t){\cal{O}}_{R_1}(t')\rangle$ by virtue of the $r\to-r$ symmetry in the metric and EOMs.

The correlator between boundaries $\langle{\cal{O}}_{R_1}(t){\cal{O}}_{R_2}(t')\rangle$ requires a bulk solution with sources on both sides turned on, i.e.
\be
\chi(t,r)=\int dt'\int_{{\cal{F}}}\frac{d\omega}{2\pi}e^{-i\omega(t-t')}(q_{1,\omega}(-r)\chi_L(t')+q_{1,\omega}(r)\chi_R(t'))
\ee
whose on shell action results, see \eqref{App:S0RR},
\be 
I^0_{R_{1}+R_{2}}=I^0_{11}-i\int dt\,dt'\chi_{R_{1}}(t)K_{12}(t,t')\chi_{R_{2}}(t')+I^0_{22}
\ee
leading to
\begin{align} \label{AdS2G-RL}
\langle{\cal{O}}_{R_1}(t){\cal{O}}_{R_2}(t')\rangle  =-i K_{12}(t,t')
  = \frac{ 2^{-\Delta } \Gamma (\Delta +1)}{\sqrt{\pi } \Gamma \left(\Delta
   -\frac{1}{2}\right)} \frac{1 }{\left(\cos((t-t')(1-i\epsilon))+1\right)^{\Delta}}\;.
\end{align}
that correctly reproduces a time ordered correlator with no singularities at $t=t'$.

\subsection{Expectation values}

We now compute expectation values of the KG probe field $\chi$ for general initial and final holographic excited states introduces in Sec. \ref{Sec:HES}.
To do this we can reduce the problem to an pure Euclidean set-up and look for the configuration at the moment of time reflection symmetry $t=\tau=0$. 
The Euclidean metric is obtained by a Wick rotation $t\to-i\tau$ on \eqref{App:Lmetric}, i.e.
\begin{equation}\label{App:Emetric}
    ds^2=(r^2+1)d\tau^2+\frac{dr^2}{r^2+1}, \qquad r,\tau\in(-\infty,\infty)\;
\end{equation}
Notice that this foliation suggests the existence of two disconnected boundaries in the pure Euclidean AdS$_2$ scenario which is not true, see Fig. \ref{Fig:Fol}. We insist on this foliation here for the two boundaries will become actually disconnected in the Thermal AdS$_2$ scenario which is the one relevant for JT/SYK holography and covered in Sec. \ref{Thermal}.

As explained in Sec. \ref{Sec:HES}, to describe an initial holographic excited state, we should put a general Euclidean source in the Euclidean past region $\tau<0$. Our coordinates force us to further split this into 2 pieces that we can call $\chi_{I_1}$ and $\chi_{I_2}$, in the sense that they both belong to the same past Euclidean segment, $\chi_{I_{1/2}}(\tau)=0$ for $\tau>0$, but are defined only on $r\to\pm\infty$ respectively.
We will show the positive energy modes that arise from sources turned on only on the past Euclidean region. Negative energy modes arise from sources put in the Euclidean future and these can be obtained by $\tau\to-\tau$ reflection symmetry. We refer to \cite{us1} for a detailed description of this construction.

The Euclidean problem analogous to \eqref{Global-EDM} on the metric \eqref{App:Emetric} with sources in the Euclidean past can be written as,
\begin{align}
    \chi_I(\tau,r)&=\int_{-\infty}^0d\tau' \int_{\cal F} \frac{d\omega}{2\pi}e^{-i\omega (-i\tau +i\tau')} \left(\chi_{I_1}(\tau')q_{1,\omega}(r)+\chi_{I_2}(\tau')q_{1,\omega}(-r)\right) \\
    &= \int_{\cal F} \frac{d\omega}{2\pi}e^{-\omega \tau} \left(\bar\chi_{I_1}(\omega)q_{1,\omega}(r)+\bar\chi_{I_2}(\omega)q_{1,\omega}(-r)\right) \;.\label{App:VEVSol}
\end{align}
We have written the solution in terms of the Lorentzian solutions found above for convenience.
The $\bar\chi_{I_{1/2}}(\omega)$ stand for the Laplace transform of the asymptotic sources. To be precise, the solution meets
\begin{align}
    \chi_I(\tau,r)&\sim r^{\Delta-1} \chi_{I_1}(\tau) \qquad\qquad r\to+\infty\\
    \chi_I(\tau,r)&\sim r^{\Delta-1} \chi_{I_2}(\tau) \qquad\qquad r\to-\infty\;.
\end{align}

The physical propagating modes can be read by considering first $\tau\sim0$ but $\tau>\tau'$, i.e. we stand in the Euclidean future of all asymptotic sources, such that one can use the Residue theorem on the $\omega$ integral. One then takes $\tau\to it$, resulting in
\begin{align}
    \chi_I( t,r)&=\sum_{n=0}^\infty e^{-i(\Delta+ n) t} \left(\bar\chi_{I_1}(\Delta + n ) + \bar\chi_{I_2}(\Delta + n)\right) \mathop{\mathrm{Res}}_{\omega=\Delta+n}q_{1,\omega}(r)
\end{align}
Since we have only excited the Euclidean past, both integrals closed upwards and only positive energy modes $\omega=+(\Delta+n)$, $n\geq0$ appear. 
The negative frequency modes $\omega=-(\Delta+n)$ should appear when sources on the Euclidean future $\tau'>0$ are turned on. The corresponding solution yields
\begin{align}
\chi_F( t,r)&=\sum_{n=0}^\infty e^{+i(\Delta+ n) t} \left(\bar\chi_{F_1}(\Delta + n ) + \bar\chi_{F_2}(\Delta + n)\right) \mathop{\mathrm{Res}}_{\omega=-(\Delta+n)}q_{1,\omega}(r)
\end{align}

Notice that both $\chi_I(t,r)$ and $\chi_F(t,r)$ solutions consist of N modes \eqref{App:NModes}.
The expectation value of boundary operators ${\cal O}_{R/L}$ can be obtained via the BDHM dictionary, see \eqref{BDHM2-ave}, see also \cite{us1}
\begin{align}\label{AdS2G-R}
    \langle \chi_F | {\cal O}_R(t)| \chi_I\rangle 
    &\equiv (2\Delta-1)\lim_{r\to+\infty} r^{\Delta} \left(\chi_I( t,r)+\chi_F( t,r)\right) \\\nn
    &=\sum_{n=0}^\infty \left(
    e^{-i(\Delta+ n) t} \left(\bar\chi_{I_1}(\Delta + n ) + \bar\chi_{I_2}(\Delta + n)\right)
    +e^{i(\Delta+ n) t} \left(\bar\chi_{F_1}(\Delta + n ) + \bar\chi_{F_2}(\Delta + n)\right) \right) F_{\Delta,n}
    \end{align}
    \begin{equation}\label{AdS2G-L}
        \langle \chi_F | {\cal O}_L(t)| \chi_I\rangle \equiv (2\Delta-1)\lim_{r\to-\infty} r^{\Delta} \left(\chi_I( t,r)+\chi_F( t,r)\right) =\langle \chi_F | {\cal O}_R(t)| \chi_I\rangle
    \end{equation}
The inherited normalization from AdS is,
\begin{align}
   F_{\Delta,n}&= 
   \frac{ \Gamma \left(\frac{3}{2}-\Delta \right)
   \Gamma (n+2 \Delta )}{4^{\Delta-1 }\pi  \Gamma \left(\Delta
   -\frac{1}{2}\right) \Gamma (n+1)}
\end{align}
As a final comment, it has been noticed in the literature that this inherited operator normalization coming from holography is consistent between the GKPW and BDHM \cite{Harlow:2011ke,us1,us4}. This amounts to a non-trivial match since the GKPW prescription relies only on semiclassical bulk computations and the BDHM prescription requires to build an orthonormal set of modes to canonically quantize the fields. Such an identity was proven useful to analytically compute the correct normalization for these orthonormal modes \cite{us3} as well as to compute Bogoliubov coefficients between different quantizations \cite{Belin}.

\end{document}